%% file: main.tex
\documentclass[10pt, conference, letterpaper]{IEEEtran}
\pdfoutput=1
\usepackage{algorithm}
\usepackage{subcaption}
\usepackage{algpseudocode}
\usepackage{amsmath}
\usepackage{balance}
\usepackage{cite}
\usepackage{enumitem}
\usepackage{epsfig}
\usepackage{epstopdf}
\usepackage{float}
\usepackage{graphics}
\usepackage{graphicx}
\usepackage{hhline}
\usepackage{multirow}
\usepackage{multicol}
\usepackage{supertabular}
\usepackage{cuted}
\usepackage{mathtools}
\usepackage[table]{xcolor}
\usepackage{url}
\usepackage{amssymb}
\usepackage{pifont}

\IEEEoverridecommandlockouts
\begin{document}

\def\sharedaffiliation{%
\end{tabular}
\begin{tabular}{c}}
\newcommand{\ud}{\mathrm{d}}
\newcommand*{\TitleFont}{%
      \usefont{\encodingdefault}{\rmdefault}{b}{n}%
      \fontsize{17}{20}%
      \selectfont}
\title{Next Generation Wi-Fi and 5G NR-U in the 6~GHz Bands: Opportunities \& Challenges}
\author{
Gaurang Naik$^*$, Jung-Min (Jerry) Park$^*$, Jonathan Ashdown$^\dag$, William Lehr$^\ddag$ \\
$^*$Bradley Department of Electrical and Computer Engineering, Virginia Tech, Arlington, VA, USA\\
$^\dag$Department of Electrical \& Computer Engineering, 
SUNY Polytechnic Institute, Utica, NY, USA \\
$^\ddag$Computer Science and Artificial Intelligence Laboratory, Massachusetts Institute of Technology, Boston, MA, USA \\
\{gaurang, jungmin\}@vt.edu, Jonathan.Ashdown@sunypoly.edu, wlehr@mit.edu
}  

\maketitle

\begin{abstract}
The ever-increasing demand for unlicensed spectrum has prompted regulators in the US and Europe to consider opening up the 6~GHz bands for unlicensed access. These bands will open up 1.2~GHz of additional spectrum for unlicensed radio access technologies (RATs), such as Wi-Fi and 5G New Radio Unlicensed (NR-U), in the US and if permitted, 500~MHz of additional spectrum in Europe. The abundance of spectrum in these bands creates new opportunities for the design of mechanisms and features that can support the emerging bandwidth-intensive and latency-sensitive applications. However, coexistence of unlicensed devices both with the bands' incumbent users and across different unlicensed RATs present significant challenges. In this paper, we provide a comprehensive survey of the existing literature on various issues surrounding the operations of unlicensed RATs in the 6~GHz bands. In particular, we discuss how key features in next-generation Wi-Fi are being designed to leverage these additional unlicensed bands. We also shed light on the foreseeable challenges that designers of unlicensed RATs might face in the near future. Our survey encompasses key research papers, contributions submitted to standardization bodies and regulatory agencies, and documents presented at various other venues. Finally, we highlight a few key research problems that are likely to arise due to unlicensed operations in the 6~GHz bands. Tackling these research challenges effectively will be critical in ensuring that the new unlicensed bands are efficiently utilized while guaranteeing the interference-free operation of the bands' incumbent users. 
\end{abstract}

\input{introduction.tex}
\input{related.tex}
\input{background.tex}
\input{opportunities.tex}
\input{co-channel-incumbent.tex}
\input{co-channel-unlicensed.tex}
\input{adj-channel.tex}

\input{challenges.tex}

\input{conclusions.tex}

\bibliographystyle{ieeetr}
\bibliography{main}

\end{document}

%% file: introduction.tex
\section{Introduction}
\label{sec:introduction}

\begin{figure*}[htb]
    \centering
    \includegraphics[trim={4cm 0cm 10.5cm 0cm},clip,scale=0.65,angle=-90]{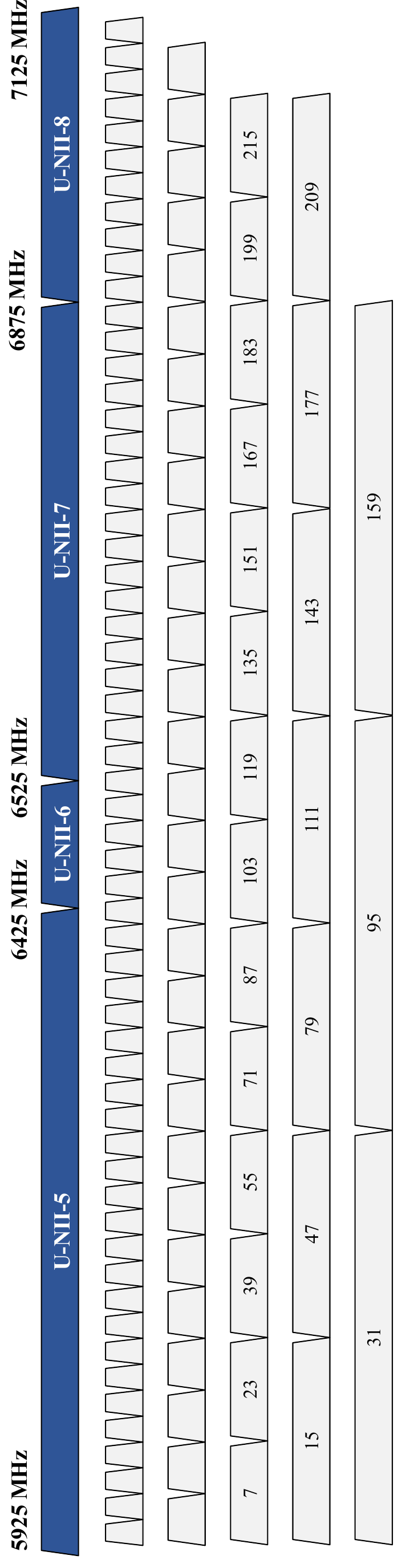}
    \caption{The 6~GHz channels for unlicensed access in the US.}
    \label{fig:6GHz-channels-us}
\end{figure*}

\begin{figure}[htb]
    \centering
    \includegraphics[trim={12cm 0cm 3cm 15cm},clip,scale=0.65,angle=-90]{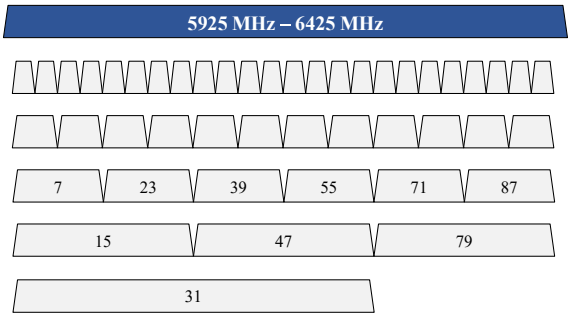}
    \caption{The 6~GHz channels for unlicensed access in Europe.}
    \label{fig:6GHz-channels-europe}
\end{figure}

Wireless devices operating in unlicensed bands have become an integral part of our lives today. The Federal Communications Commission (FCC) in the US first opened up the 2.4-2.4835~GHz and 5.725-5.85~GHz bands for unlicensed access in 1985~\cite{6ghz-nprm}. Since then, several unlicensed radio access technologies (RATs)---most notably IEEE 802.11-based Wi-Fi, Bluetooth, and ZigBee---have been developed that can not only operate in these bands but also coexist with each other. The 2.4~GHz band, referred to as the Industrial, Scientific, and Medical (ISM) band, is open for unlicensed access worldwide, while unlicensed RATs such as Wi-Fi are allowed to operate in several portions of the 5~GHz bands in most regions across the world~\cite{naik2018coexistence}. In the US, these 5~GHz bands are referred to as the Unlicensed National Information Infrastructure (U-NII) bands. 

Wi-Fi is arguably the most popular unlicensed RAT that provides mobile and high-speed Internet access over wireless local area networks (WLANs). Wi-Fi devices are ubiquitous in today's home and enterprise wireless networks, with an estimated 9.5 billion devices in use~\cite{wifi-alliance}. Furthermore, with the growing popularity of emerging applications such as wireless augmented reality (AR), virtual reality (VR), and mobile gaming, the demand for Wi-Fi-based high-throughput, high-reliability and low-latency connectivity is rapidly increasing. Consequently, unlicensed wireless spectrum is a resource that is sought after more than ever before. To cater to this growing need for unlicensed spectrum, FCC in the US and the European Commission (EC) in Europe initiated studies to determine the feasibility of unlicensed operations in the 6~GHz bands. Specifically, the EC mandated a feasibility study of unlicensed operations in the 5.925-6.425~GHz band in Europe~\cite{ecc-first}. At the same time, the FCC Notice of Proposed Rulemaking (NPRM)~\cite{6ghz-nprm} sought comments on opening up the 5.925-7.125~GHz band for unlicensed access in the US. In the US, this band will be divided into four sub-bands: U-NII-5 (5.925--6.425~GHz), U-NII-6 (6.425--6.525~GHz), U-NII-7 (6.525--6.875~GHz), and U-NII-8 (6.875--7.125~GHz). Spectrum sharing rules for the U-NII-5 through U-NII-8 bands in the US were recently finalized by the FCC in the 6~GHz Report \& Order (R\&O)~\cite{fcc-ro}. 

Unlicensed channels in the 6~GHz bands in the US and Europe are shown in Fig.~\ref{fig:6GHz-channels-us} and Fig.~\ref{fig:6GHz-channels-europe}, respectively.

The 6~GHz bands are deemed critical in supporting emerging wireless AR/VR and mobile gaming applications, which are characterized by stringent Quality of Service (QoS) requirements~\cite{11-18-2009r6}. The additional spectrum that the 6~GHz bands will unlock (500~MHz in Europe and 1.2~GHz in the US) will significantly increase the amount of unlicensed spectrum available in these regions. For example, the 6~GHz bands will more than double the current amount of unlicensed spectrum in the US. The peak data rates achievable using such large amounts of unlicensed spectrum can rival those achieved in the millimeter wave (mmWave) bands. However, devices using the 6~GHz bands can achieve these data rates without the challenges encountered in the mmWave bands (such as high propagation losses, sensitivity to blockage, etc.~\cite{niu2015survey}). Further, the abundant unlicensed spectrum in the 6~GHz bands is promising in terms of enabling features and mechanisms that can support QoS-sensitive applications. This has prompted industry stakeholders to expedite their efforts on research and development of new mechanisms that leverage the abundant spectrum available in the 6~GHz bands. In the development of IEEE 802.11be~\cite{wifi-7}---the successor to the upcoming IEEE 802.11ax~\cite{wifi-6} standard---one of the critical goals is to effectively utilize the 6~GHz bands. Furthermore, the availability of the first set of Wi-Fi devices capable of operating in the 6~GHz bands, which will be based on IEEE 802.11ax, has already been announced~\cite{qualcomm-announcement, broadcom-6e}.

The 5G New Radio Unlicensed (NR-U) is another RAT that is designed to operate in the 6~GHz bands alongside Wi-Fi~\cite{tr38889, RP-182878, tr37890}. Spectrum sharing between next generation Wi-Fi and NR-U is, thus, imminent in the 6~GHz bands. NR-U is in the final stages of its development as a Third Generation Partnership Project (3GPP) work item, and its specifications are set to be out in 3GPP Release 16~\cite{toskala20205}. NR-U is a successor to 3GPP's Release 13/14 Long Term Evolution (LTE) License Assisted Access (LAA)~\cite{kwon2016licensed}, and derives its physical (PHY) layer from the 5G NR~\cite{ahmadi}. Coexistence between LTE-LAA and Wi-Fi was extensively studied for the 5~GHz bands~\cite{chen2016coexistence, naik2018coexistence, ahmadi}. However, a significant constraint placed on the design of LTE-LAA was that LTE-LAA devices must coexist fairly with a large number of already deployed Wi-Fi devices. Consequently, the channel access protocols used in LTE-LAA were forced to align with those used by Wi-Fi devices. However, such constraints do not apply in the 6~GHz bands, where no unlicensed devices currently operate, and both Wi-Fi and NR-U devices will operate on a secondary basis for the first time~\cite{11-19-1083r1}. 

The first two RATs capable of operating in the 6~GHz bands, i.e., IEEE 802.11ax and NR-U, are both undergoing final stages of development in their respective standardization bodies. This, coupled with the fact that the \textit{greenfield} 6~GHz spectrum provides an opportunity to design novel technology-neutral coexistence mechanisms~\cite{11-19-1083r1}, has fueled substantial interest in the investigation of new approaches to coexistence. Toward this goal, several organizations are actively studying issues related to the coexistence of Wi-Fi and NR-U with a focus on the 6~GHz bands. These organizations include the IEEE 802.11 Coexistence Standing Committee (SC) and the European Telecommunications Standards Institute (ETSI) Broadband Radio Access Network (BRAN).

In addition to coexistence among unlicensed RATs, NR-U and Wi-Fi must both coexist with the incumbent users of the 6~GHz bands. The FCC R\&O mandates the use of an Automatic Frequency Coordination (AFC) system in the U-NII-5 and U-NII-7 bands, whereby unlicensed devices must first contact the AFC system database and acquire a list of permissible frequencies. For the U-NII-6 and U-NII-8 bands, unlicensed devices shall be allowed to operate only in indoor settings at low transmission powers, thereby relying on building entry losses to mitigate interference at incumbent receivers~\cite{fcc-ro}. While these mechanisms seem suitable at enabling harmonious coexistence between the incumbents and unlicensed devices, several concerns have been raised by the incumbent users. The resolution of these issues is critical to ensure that the incumbent users' performance, which often have stringent outage budgets~\cite{ex-parte-fwcc}, remains unaffected.

Table~\ref{tab:Summary} contains the summary of 6~GHz specific operations, as permitted by the FCC, in the US.

\begin{table*}[htb]
    \centering
    \begin{tabular}{|p{2.5cm}|p{1.4cm}|p{1.4cm}|p{1.4cm}|p{1.4cm}|p{1.4cm}|p{1.4cm}|p{1.4cm}|p{1.4cm}|}
    \hline
       & \multicolumn{2}{|c|}{\textbf{U-NII-5}} &      \multicolumn{2}{|c|}{\textbf{U-NII-6}} & \multicolumn{2}{|c|}{\textbf{U-NII-7}} & \multicolumn{2}{|c|}{\textbf{U-NII-8}} \\ \hline
       \textbf{Frequency range} & \multicolumn{2}{|p{3cm}}{5.925--6.425~GHz} & \multicolumn{2}{|p{3cm}}{6.425--6.525~GHz} & \multicolumn{2}{|p{3cm}}{6.525--6.875~GHz} & \multicolumn{2}{|p{3cm}|}{6.875--7.125~GHz} \\ \hline
       \multirow{4}{*}{\textbf{No. of channels}} & \textbf{20~MHz} & 24 & \textbf{20~MHz} & 4 & \textbf{20~MHz} & 18 & \textbf{20~MHz} & 11 \\ \cline{2-9}
       & \textbf{40~MHz} & 12 & \textbf{40~MHz} & 1 & \textbf{40~MHz} & 9 & \textbf{40~MHz} & 5 \\ \cline{2-9}
       & \textbf{80~MHz} & 6 & \textbf{80~MHz} & 0 & \textbf{80~MHz} & 4 & \textbf{80~MHz} & 2 \\ \cline{2-9}
       & \textbf{160~MHz} & 3 & \textbf{160~MHz} & 0 & \textbf{160~MHz} & 2 & \textbf{160~MHz} & 1\\ \hline
       \textbf{Regulatory constraints} & \multicolumn{2}{|p{3cm}|}{Indoor operations permitted. Outdoor operations permitted \emph{only if} device is outside exclusion zones.} & \multicolumn{2}{|p{3cm}|}{Indoor operations permitted. Outdoor operations \emph{not} permitted.} & \multicolumn{2}{|p{3cm}|}{Indoor operations permitted. Outdoor operations permitted \emph{only if} device is outside exclusion zones.} & \multicolumn{2}{|p{3cm}|}{Indoor operations permitted. Outdoor operations \emph{not} permitted.} \\ \hline
        \multirow{4}{*}{\textbf{Outdoor constraints}} & \multicolumn{2}{|p{3cm}|}{Devices must connect to the AFC system database.} & \multicolumn{2}{|c|}{N/A} & \multicolumn{2}{|p{3cm}|}{Devices must connect to the AFC system database.} & \multicolumn{2}{|c|}{N/A} \\ 
       & \multicolumn{2}{|p{3cm}|}{Max power: 36dBm (AP)/ 30dBm (client)} & \multicolumn{2}{|p{3cm}|}{} & \multicolumn{2}{|p{3cm}|}{Max power: 36dBm (AP)/ 30dBm (client)} & \multicolumn{2}{|p{3cm}|}{} \\ \hline
       \multirow{4}{*}{\textbf{Indoor constraints}} & \multicolumn{8}{|p{12cm}|}{- Devices \textit{cannot} be weather resistant. }  \\ 
        & \multicolumn{8}{|p{12cm}|}{- Devices \textit{cannot} be equipped with external antennas.}  \\ 
        & \multicolumn{8}{|p{12cm}|}{- Devices \textit{cannot} be operated on battery power.}  \\ 
        & \multicolumn{8}{|p{12cm}|}{- Max power: 30dBm (AP)/ 24dBm (client)} \\ \hline
    \end{tabular}
    \caption{Summary of 6~GHz operations in the US.}
    \label{tab:Summary}
\end{table*}

In this paper, we present findings from our comprehensive survey of key literature available on the issues mentioned above. The main contributions of this paper are as follows. 

\begin{itemize}
    \item We provide a comprehensive overview of the opportunities unlocked by the opening up of the 6~GHz spectrum for unlicensed access, and the challenges most likely to be faced in the adoption of these bands by unlicensed services. Our survey includes findings from research papers, contributions made to regulatory agencies and standardization bodies, including IEEE and 3GPP, and documents and presentations delivered at other fora. 
    \item Based on the discussions presented in the surveyed literature, we identify key research problems that need to be addressed to ensure harmonious coexistence between the incumbents and the heterogeneous RATs that will operate in the 6 GHz bands.
\end{itemize}

Although discussions presented in this paper reflect the current state of affairs at various standardization bodies and regulatory agencies, several aspects of 6~GHz operations are still to be finalized. Thus, discussions presented in this paper, including features, mechanisms, coexistence solutions, and other operational issues, are subject to change by the time all 6~GHz operational rules are finalized. 

A summary of acronyms used in this paper is provided in Table~\ref{tab:my_label}. The remainder of this paper is outlined as follows. Sec.~\ref{sec:related} reviews the related literature. Sec.~\ref{sec:background} provides a brief background of the stakeholder technologies, which include incumbent users, unlicensed RATs likely to operate in the 6~GHz bands, and critical services that operate in adjacent bands. Next, in Sec.~\ref{sec:opportunities} we present findings from our survey on the opportunities presented by the opening up of the 6~GHz bands for unlicensed operations. Sec.~\ref{sec:incumbent} then discusses issues surrounding the coexistence of unlicensed devices with incumbent users of the 6~GHz bands, while Sec.~\ref{sec:unlicensed} provides details on coexistence between NR-U and Wi-Fi. Sec.~\ref{sec:adjacent} highlights the interference originating from unlicensed devices (operating in the lower end of the 6~GHz bands) at incumbent users of the 5.9~GHz band. In Sec.~\ref{sec:challenges}, we outline key research challenges that we have identified during our survey. Finally, Sec.~\ref{sec:conclusions} concludes the paper. 


\begin{table}[htb]
    \centering
    \begin{tabular}{|l|l|}
    \hline
        \textbf{Acronym} & \textbf{Full name} \\ \hline \hline
        3GPP & Third Generation Partnership Project \\ 
        ACK & Acknowledgement \\ 
        AFC & Automatic Frequency Coordination \\ 
        AP & Access Point \\ 
        AR & Augmented Reality \\ 
        BRAN & Broadband Access Radio Networks \\
        CBTC & Communication based train control \\ 
        CSMA/CA & Carrier Sense Multiple Access with Collision Avoidance \\ 
        C-V2X & Cellular Vehicle-to-Everything \\ 
        CW & Contention Window \\ 
        DFS & Dynamic Frequency Selection \\ 
        DSRC & Dedicated Short Range Communications \\ 
        EC & European Commission \\ 
        ECC & European Communications Committee \\ 
        ED & Energy Detection \\ 
        EHT & Extremely High Throughput \\ 
        EIRP & Effective Isotropic Radiated Power \\ 
        ETSI & European Telecommunications Standards Institute \\ 
        FBE & Frame Based Equipment \\ 
        FCC & Federal Communications Commission \\ 
        FNPRM & Further Notice of Proposed Rulemaking \\
        IEEE &  Institute of Electrical and Electronics Engineers \\ 
        IFS & Interframe spacing \\ 
        ISM & Industrial, Scientific \& Medical (band) \\ 
        ITS & Intelligent Transportation Systems \\ 
        LAA & License Assisted Access \\ 
        LBE & Load Based Equipment \\ 
        LBT & Listen Before Talk \\
        LTE & Long Term Evolution \\ 
        LPI & Low Power Indoor \\ 
        MAC & Medium Access Control (layer) \\ 
        MCS & Modulation and Coding Scheme \\ 
        MIMO & Multiple Input Multiple Output \\ 
        MLA & Multi-Link Aggregation \\ 
        MU & Multi-User \\ 
        NPRM & Notice of Proposed Rulemaking \\ 
        NR & New Radio \\ 
        NR-U & New Radio Unlicensed \\ 
        OFDMA & Orthogonal Frequency Division Multiple Access \\ 
        PHY & Physical (layer) \\ 
        PIFS & Point Coordination Function Interframe Spacing \\ 
        QAM & Quadrature Amplitude Modulation \\ 
        QoS & Quality of Service \\ 
        R\&O & Report and Order \\
        RAT & Radio Access Technology \\
        RU & Resource Unit \\ 
        SINR & Signal-to-Interference-plus-Noise Ratio \\ 
        STA & Station \\ 
        STR & Simultaneous transmit-receive \\ 
        TXOP & Transmission opportunity \\ 
        UE & User Equipment \\ 
        U-NII & Unlicensed National Information Infrastructure \\
        UWB & Ultra wideband \\
        V2X & Vehicle-to-Everything \\ 
        VLP & Very Low Power \\ 
        VR & Virtual Reality \\ 
        WLAN & Wireless Local Area Network \\ \hline
    \end{tabular}
    \caption{Summary of acronyms used in this paper.}
    \label{tab:my_label}
\end{table}

%% file: related.tex
\section{Related Work}
\label{sec:related}

Unlicensed operations in the 6~GHz bands have been under consideration since 2018 and have been finalized in the US recently. Consequently, related literature on the opportunities presented by the 6~GHz unlicensed bands and the challenges likely to be encountered is scarce. At present, the information on 6~GHz rules and regulations is scattered across several documents, where the primary sources include the FCC R\&O~\cite{fcc-ro}, reports of studies conducted in Europe~\cite{ecc-report-302,cept-report-73}, and stakeholders' contributions to regulatory agencies. On the other hand, technologies that are the most likely to operate in these new unlicensed bands have been garnering considerable attention in the recent years. In this section, we briefly summarize the related literature on these technologies.

The first generation of Wi-Fi devices that will likely operate in the 6~GHz bands will be based on the IEEE 802.11ax standard. In the recent years, researchers across the industry and academia have investigated several aspects related to IEEE 802.11ax. These include the use of MU-OFDMA~\cite{bellalta2019ap}, spatial reuse~\cite{afaqui2015evaluation}, Target Wake Time~\cite{nurchis2019target}, etc. Reference~\cite{khorov2018tutorial} provides an excellent survey on these topics in relation to IEEE 802.11ax. As the specifications of IEEE 802.11ax are close to completion, the work on its successor, IEEE 802.11be, is actively ongoing. Even though features and mechanisms that will eventually constitute IEEE 802.11be are being debated, a few recent papers provide a very comprehensive and in-depth review on IEEE 802.11be. The most notable ones include references~\cite{khorov2020current} and~\cite{lopez2019ieee}.

On the cellular side, 5G NR-U will be developed atop the PHY layer of the 5G NR. References~\cite{ahmadi} and~\cite{dahlman20185g} describe the 5G NR, including it's PHY layer, architectural aspects, operating scenarios, etc., in comprehensive detail. Furthermore, references~\cite{lien20175g, lin20195g} provide an excellent summary on the most notable and salient features of 5G NR. Although the inheritance of the 5G NR PHY layer allows for the smooth integration of NR-U with NR-based cellular networks, it's MAC layer must be designed carefully to coexist with other unlicensed RATs. Thus, the MAC layer of NR-U and it's associated protocols are derived from the predecessor unlicensed RAT of NR-U, i.e., LTE-LAA~\cite{kwon2016licensed}. During the design of LTE-LAA, one of the most critical considerations was its fair coexistence with Wi-Fi. This subject is a thoroughly investigated one, and references~\cite{chen2016coexistence, naik2018coexistence} provide a comprehensive survey on the related literature. 

Our paper distinguishes itself from the above works in several key ways. The central focus of this paper is the 6~GHz unlicensed bands---a subject not addressed in any of the aforementioned references. We elaborate on how these bands can be effectively utilized in catering to the growing need for unlicensed spectrum (Sec.~\ref{sec:opportunities}) and the challenges that are likely to be encountered in meeting these needs. In doing so, we introduce and elaborate on key features and mechanisms that are most relevant to unlicensed operations in the 6~GHz bands (Sec.~\ref{subsec:unlicensed}). Although some of these features have been thoroughly discussed across the aforementioned references, our paper presents them in the context of their importance in the 6~GHz bands. The challenges that we discuss include coexistence of unlicensed RATs with the incumbent users of the bands (Sec.~\ref{sec:incumbent}), across different RATs (Sec.~\ref{sec:unlicensed}), and interference at incumbent users of the adjacent band (Sec.~\ref{sec:adjacent}). Furthermore, the gravity of coexistence issues arising with the incumbent services in the 6~GHz bands can be appreciated only through an understanding of the nature of the incumbent services operating in these bands, and their typical use-cases. Since our paper is, to the best of our knowledge, the first to address the 6~GHz unlicensed bands, a discussion on the incumbent services in the 6~GHz bands is missing in the literature (which we bridge in Sec.~\ref{subsec:incumbent}).

%% file: background.tex
\section{Background}
\label{sec:background}

\subsection{Incumbent Technologies}
\label{subsec:incumbent}

As noted in Sec.~\ref{sec:introduction}, the FCC regulations in the US divide the 6~GHz bands into four sub-bands, i.e., U-NII-5 through U-NII-8 as shwon in Fig.~\ref{fig:6GHz-channels-us}. These sub-bands will extend the U-NII regime from the 5~GHz bands to the 6~GHz bands. Unlicensed devices operating in these bands must not interfere with services provided by its incumbent users. The prominent incumbent users of the 6~GHz bands are described below.

The 6~GHz bands are home to several non-federal incumbent users in the US~\cite{fccUnii5}. Among these, the most prominent ones are the fixed point-to-point services~\cite{6msc}. These services are used for providing highly-reliable backhaul links to critical services such as police and fire vehicles, electric grids, coordination of train movements, etc. A majority of these fixed point-to-point services are licensed to operate in the U-NII-5 and U-NII-7 bands~\cite{fccUnii5}. Safety-related fixed service links in these two bands have a reliability requirement of 99.9999\%, while most other links have a reliability requirement of 99.999\%~\cite{ex-parte-fwcc-indoor}. Such point-to-point microwave links are also used for providing backhaul to cellular mobile networks, e.g., links between the eNodeB/gNodeB of an LTE/5G network and its core network. Services provided using the U-NII-5 and U-NII-7 bands are such that the locations of incumbent users can be known beforehand and do not change frequently. Furthermore, new incumbent users in these bands are added infrequently~\cite{fcc-ro}. Consequently, the FCC R\&O mandates the use of the AFC system in these two bands for coexistence between unlicensed and incumbent users. Unlicensed devices that operate in these bands must first query the AFC database and obtain a list of permissible frequencies. We discuss issues specific to the AFC database-driven coexistence in Sec.~\ref{sec:incumbent}.

The U-NII-6 and U-NII-8 bands, on the other hand, are licensed to users whose locations cannot be determined accurately at all times. These users include local television transmission services and cable television relay services~\cite{fccUnii5}. The former services are used by television pickup stations to stream content from special events or remote locations  (e.g., electronic newsgathering services) to central locations such as the television studio. The latter services operate similarly in that fixed/mobile transmitting stations send audio and video data back to the receiving stations. Incumbent users in these bands are \textit{nomadic} in that the locations of transmitters and receivers can change frequently based on the location of special events such as sporting events and concerts. Consequently, licenses are granted to these users to operate in vast areas and throughout the frequency bands for maximum flexibility. This, however, implies that a geo-location database-based coexistence approach cannot be used to guarantee the presence or absence of such incumbent users at any given location and is, therefore, ineffective. 

In addition to the aforementioned incumbent users, fixed satellite services are also offered using various portions of the 6~GHz bands. For example, the U-NII-5 band supports earth-to-space fixed satellite services that cater to applications like content distribution for radio and television broadcasters. However, since receivers of these applications (i.e., satellites) are located far from potential unlicensed interferers, individual unlicensed devices are likely to cause no interference at satellite receivers. Thus, the FCC R\&O does not require unlicensed devices to take any explicit measures in order to coexist with earth-to-space fixed satellite services~\cite{fcc-ro}. 

The U-NII-6 band also enables applications that use ultra wideband (UWB) systems. These systems operate under FCC's Part 15 rules for unlicensed operations. They include real-time locating systems, which are used for tool tracking and worker safety in industrial environments, ball and player tracking in National Football League matches~\cite{ex-parte-zebra}, airport baggage handling systems, and robotic applications~\cite{ex-parte-uwb}. UWB systems are characterized by extremely low transmit powers (less than -41.3~dBm/MHz~\cite{fcc-ro}) spread out over large bandwidths, which makes these systems extremely susceptible to external interference. 

In Europe, the 5.925-6.425~GHz band---which has the same frequency range as the U-NII-5 band in the US---is under consideration for unlicensed use by radio local area networks, including Wi-Fi. The incumbent users of this band include fixed services and fixed satellite (earth-to-space) services~\cite{ecc-report-302}. This band also supports long-distance point-to-point links that are used to backhaul mobile broadband networks similar to that in the US~\cite{ecc-first}. Some countries in Europe also use parts of this band for railroad train control, referred to as the communication based train control (CBTC)\footnote{CBTC systems use the 5.915--5.935~GHz band in France, 5.925--5.975~GHz band in Denmark, and 5.905--5.925~GHz band in Spain~\cite{ecc-report-302}.}~\cite{ecc-first}. Additionally, like in the US, this band supports unlicensed users that use UWB systems. Because UWB systems are unlicensed users, they operate on a non-interference and non-protected basis. Note that with a few exceptions, the nature of the incumbent users in Europe is the same as those in the US. Thus, issues that arise due to unlicensed operations in the 6~GHz bands are likely to be similar in the two regions. 

\subsection{Unlicensed Technologies}
\label{subsec:unlicensed}

\subsubsection{Wi-Fi}
\label{subsubsec:wi-fi}

Wi-Fi refers to the WLAN technology that is based on the IEEE 802.11 specifications for its PHY and medium access control (MAC) layers. At present, Wi-Fi devices mainly operate in unlicensed portions of the 2.4~GHz ISM and 5~GHz bands~\cite{naik2018coexistence}. Since 1999, IEEE 802.11 specifications have continuously evolved from the IEEE 802.11b, which provided a peak PHY layer rate of 11~Mbps to the recent IEEE 802.11ac, the second wave products of which boasted a peak PHY rate of 6.77~Gbps. The recently concluded IEEE 802.11ax specifications are the latest in Wi-Fi's continuing line of evolution.

Wi-Fi 6 refers to the set of Wi-Fi devices that are certified as per the IEEE 802.11ax specifications~\cite{wifi-6}. The first set of commercial W-Fi 6 devices capable of operating in the 6~GHz bands (referred to as Wi-Fi 6E) is already available~\cite{announcement}. In contrast to the previous generation of Wi-Fi standards, where the focus was primarily on increasing the offered throughput, Wi-Fi 6 introduces a range of features that are expected to enhance the perceived experience of Wi-Fi users. These include the use of Multi-User Orthogonal Frequency Division Multiple Access (MU-OFDMA) for channel access, spatial re-use, target wake time for improved power efficiency, and the use of higher-order modulation and coding schemes (MCS) such as 1024 Quadrature Amplitude Modulation (QAM) for increased throughput~\cite{bellalta2016ieee}.

One of the most prominent features of Wi-Fi 6 is the use of MU-OFDMA for uplink and downlink transmissions~\cite{bellalta2016ieee}. Traditionally, Wi-Fi users have used Carrier Sense Multiple Access with Collision Avoidance (CSMA/CA) as the MAC layer protocol for channel access~\cite{naik2018coexistence}. Devices using CSMA/CA must transmit one at a time over the entire channel bandwidth. Wi-Fi 6 users can indeed transmit using the \textit{legacy} CSMA/CA-based contention mechanism. This is especially important because, in a majority of the current unlicensed bands, Wi-Fi 6 devices must coexist with a large number of legacy, i.e., IEEE 802.11ac and older, Wi-Fi devices. MU-OFDMA, on the other hand, allows for division of the channel into Resource Units (RUs), which can then be assigned to individual Wi-Fi users based on their traffic demands (either in the downlink or uplink). However, since Wi-Fi 6 devices need not coexist with legacy Wi-Fi devices in the 6~GHz bands, Wi-Fi 6 networks can restrict devices to use only MU-OFDMA for uplink and downlink transmissions. As a result, the MAC layer efficiency, and consequently, user throughput, in dense usage settings will improve~\cite{naik2018performance}.

Even as Wi-Fi 6 certified devices continue to be rolled out, work on the specifications of its successor---IEEE 802.11be Extremely High Throughput (EHT)---has already begun~\cite{wifi-7,khorov2020current}. The first draft standard of IEEE 802.11be is expected to be completed by May 2021~\cite{11-19-0787r2}. Target applications of IEEE 802.11be include Wi-Fi-based AR and VR, real-time gaming, and industrial automation, which are often characterized by high throughput coupled with extremely high reliability and low latency requirements~\cite{11-19-1207r4,adame2019time}. The Task Group (TG) be, i.e., TGbe, responsible for the standardization of PHY and MAC layer features that will eventually constitute IEEE 802.11be, was formed in May 2019. In contrast to IEEE 802.11ax, which was initially designed to operate in unlicensed portions of the sub-6~GHz bands, IEEE 802.11be considers the use of unlicensed bands, wherever available, in the 1~GHz to 7.125~GHz range~\cite{lopez2019ieee}. 

IEEE 802.11be will support all features introduced in 802.11ax. Additionally, 802.11be is likely to introduce new features that contribute toward meeting the high throughput, high reliability, and low latency objectives set forth by the TGbe~\cite{11-19-1095}. Some of these features include multiple Access Point (AP) coordination~\cite{11-19-0801r0,yang2019ap}, multiple input multiple output (MIMO) enhancements including provisioning support for up to 16 spatial streams~\cite{11-19-0832r0}, potential support for 4096 QAM~\cite{11-19-0637} and Hybrid Automatic Repeat Request (HARQ)~\cite{11-19-0780r0}, providing support for 240~MHz and 320~MHz channelization\footnote{IEEE 802.11ax and IEEE 802.11ac allow maximum channel bandwidth of 160~MHz.}~\cite{11-19-1889r2}, and multi-link aggregation (MLA)~\cite{11-19-0818r1}. Interested readers can refer to references~\cite{khorov2020current} and~\cite{lopez2019ieee}  for detailed descriptions of these features.  Among these novel features, two specific ones---increased channel widths and MLA---are tightly coupled to the availability of the 6~GHz bands. 

\begin{figure}
    \begin{subfigure}{0.5\textwidth}
     \centering
     \includegraphics[trim={7cm 6.5cm 11.4cm 8cm},clip,scale=0.63,angle=-90]{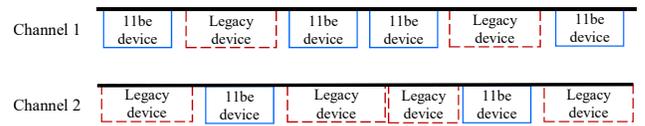}
     \subcaption{Asynchronous MLA. MLA-capable 802.11be device contends on both channels independently.}
     \label{fig:async}
    \end{subfigure}\\
    \begin{subfigure}{0.5\textwidth}
     \centering
     \includegraphics[trim={7cm 6.5cm 11.1cm 8cm},clip,scale=0.63,angle=-90]{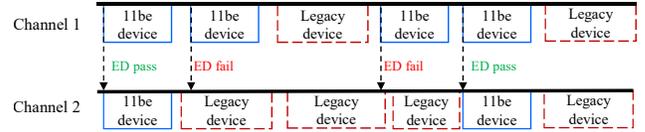}
     \caption{Synchronous single-primary MLA. MLA-capable 802.11be device contends only on channel 1.}
     \label{fig:single-sync}
    \end{subfigure}\\
   \begin{subfigure}{0.5\textwidth}
     \centering
     \includegraphics[trim={7cm 6.5cm 11.1cm 8cm},clip,scale=0.63,angle=-90]{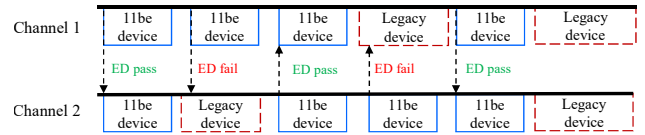}
     \caption{Synchronous multi-primary MLA. MLA-capable 802.11be device contends on both channel 1 \& 2.}
     \label{fig:multi-sync}
    \end{subfigure}
    \caption{Multi Link Aggregation Schemes in IEEE 802.11be.}
    \label{fig:mla-types}
    \vspace{-4mm}
\end{figure}

MLA will allow 802.11be devices that can operate in different bands (such as 2.4~GHz, 5~GHz, and 6~GHz) to use these bands concurrently. Using MLA, IEEE 802.11be stations (STAs) and AP can send/receive MAC service data units belonging to the same flow on multiple bands (e.g., 2.4~GHz and 5~GHz, or 5~GHz and 6~GHz) or channels (two channels in the 6~GHz band) simultaneously~\cite{11-19-0766r1}. At present, there are three MLA schemes under consideration~\cite{11-19-1144r6}. The choice of MLA scheme depends on the implementation complexity and the frequency separation between aggregated channels. These schemes are illustrated in Fig.~\ref{fig:mla-types} for aggregation of two links, although the same principles can be used for aggregation of more than two links.

Independent/asynchronous MLA is used when the two aggregated links are sufficiently far apart in frequency resulting in no or negligible inter-link cross-talk (e.g., aggregation of 2.4~GHz and 5~GHz bands, or aggregation of 2.4~GHz and 6~GHz bands). The MLA-capable AP/STA contends for channel access on both channels and transmits the packet on the contention-winning channel. Fig.~\ref{fig:async} demonstrates the operation of asynchronous MLA for two links. The name independent/asynchronous signifies that contention on the two operating channels is independent of each other. Thus, transmissions on the two channels need not be synchronized.

Synchronous/simultaneous MLA, on the other hand, is further of two types depending on whether the MLA-capable AP/STA contends on only one link (referred to as the primary) or on both links. The former scheme is referred to as synchronous/simultaneous \textit{single}-primary MLA, while the latter is referred to as synchronous/simultaneous \textit{multi}-primary MLA. In both schemes, right before the MLA-capable AP/STA wins access to the primary channel\footnote{Note that in synchronous/simultaneous multi-primary MLA, each of the two channels is a primary channel.} (i.e., when the back-off counter strikes zero), energy detection (ED) check is performed on the other channel for a duration referred to as the PIFS. PIFS stands for Point Coordination Function Interframe spacing. This other link is referred to as the secondary link for synchronous/simultaneous single primary MLA. If the ED check passes, i.e., if the energy observed on the other link is less than the ED threshold, the two channels are aggregated, and the MLA device transmits on both links. This ensures that the two channels, if aggregated, are always synchronized. Fig.~\ref{fig:single-sync} and Fig.~\ref{fig:multi-sync} demonstrate the operation of synchronous single-primary MLA and multi-primary MLA, respectively. Note that the synchronous single primary MLA bears similarities with the channel bonding feature used in 802.11n and 802.11ac~\cite{naik2018coexistence}.

Synchronous/simultaneous MLA is suitable for aggregation of links when the frequency separation between the aggregated links is small. For example, when a channel in the U-NII-5 band is aggregated with another in the U-NII-6 band, there can be significant leakage across the radio frontends. Such inter-link cross-talk can hamper the reception of signals on one link when the other link is in the transmit state. Synchronous MLA schemes maximize the likelihood of all links simultaneously being in the transmit state. However, there can still arise scenarios where only a subset of links are in the transmit state. We discuss this issue further in Sec.~\ref{sec:challenges}.

\subsubsection{5G NR-U}
\label{subsubsec:nr-u}

NR-U is a RAT that is undergoing development by the 3GPP as one of its Release 16 work items~\cite{RP-182878}. NR-U is a successor to LTE-LAA---the unlicensed flavor of LTE standardized by the 3GPP in Rel. 13---and subsequent releases. Thus, in the design of NR-U PHY and MAC layer procedures, LTE-LAA is considered as the starting point~\cite{tr38889,ahmadi}. For example, the channel access mechanism used in LTE-LAA is adopted as the baseline for NR-U operations in the 5~GHz bands. This also implies that in areas where the absence of Wi-Fi networks cannot be guaranteed, NR-U devices will operate (and hence, perform sensing) in bandwidths that are integer multiples of 20~MHz. Nevertheless, the 3GPP Technical Report 38.889~\cite{tr38889} identifies the need to customize channel access protocols according to the operating band and/or regional regulations. Thus, NR-U devices can \textit{potentially} use different channel access mechanisms in the 5~GHz and 6~GHz bands. 

NR-U derives its PHY layer from 5G NR~\cite{ahmadi,parkvall2017nr} and, thus, can benefit from leveraging PHY layer enhancements made in 5G NR. These include the use of flexible numerologies and mini-slot scheduling, among others. The use of mini-slot scheduling in NR-U is especially beneficial because sending packets in mini-slots minimizes the use of reservation signals such as those used in LTE-LAA to reserve the channel during shared channel access~\cite{zhang2017lte}. Reservation signals were required in LTE-LAA because unlicensed transmissions from LTE-LAA devices are required to be synchronized with sub-frame boundaries of the anchored licensed carrier\footnote{The spectrum used for an LTE-LAA network is used as a supplemental carrier, which is aggregated with the operator's licensed carrier---referred to as the anchor---when resources available in the licensed carrier are insufficient to support the downlink/uplink traffic.}---a constraint that came from LTE-LAA's inheritance of licensed cellular operations. This meant that LTE-LAA devices could start transmitting data and control symbols only at the beginning of the sub-frame boundary. However, LTE-LAA devices can gain access to the channel at any given instant, depending on when the channel becomes idle. Thus, reservation signals were transmitted in the time duration between gaining access to the channel and the start of the sub-frame boundary.

A critical difference between NR-U and previous 3GPP-based unlicensed RATs is that NR-U does not require a licensed primary carrier for its operation~\cite{RP-182878, kim2019new}, which was mandatory in LTE-LAA~\cite{naik2018coexistence}. In this scenario, the NR-U network is connected to 5G Core Network~\cite{RP-182878}. The ability of NR-U to operate without a licensed primary carrier is significant because this enables NR-U networks to be deployed by any parties similar to Wi-Fi AP deployments. 

The harmonious coexistence of NR-U with other unlicensed RATs necessitates the design of a MAC protocol that is efficient, yet fair toward other RATs. The channel access mechanism in NR-U will be based on Listen Before Talk (LBT)~\cite{ts37213}, much like the LBT protocol used in previous 3GPP-defined unlicensed technologies~\cite{mehrnoush2018analytical}. Based on regional regulations and/or the type of traffic transmitted by the NR-U device, there are four categories of LBT. Devices using category 1 LBT transmit immediately after a period of 16~$\mu$s once the channel becomes idle. This is typically used for acknowledgement (ACK) packets sent by the receiver upon successful reception of a packet. Category 2 LBT also involves the transmission of a packet following a fixed time interval. However, this interval is 25~$\mu$s in the case of category 2 LBT. On the other hand, devices using category 3 and 4 LBT transmit their packets after a fixed time interval (16~$\mu$s) followed by a random number of time slots (each of duration 9~$\mu$s), where this random number is picked between 0 and CW-1, CW being the parameter Contention Window. What differentiates category 3 and 4 LBT is that the CW size remains fixed in the former while it varies in the latter. Category 4 LBT is similar to the CSMA/CA protocol used in Wi-Fi and is discussed further in Sec.~\ref{sec:unlicensed}.

The 3GPP Release 16 will continue to provide support for LTE-LAA and its successors, which will continue to operate in existing 5~GHz unlicensed bands. However, the sole 3GPP-based RAT that will operate in the 6~GHz bands is NR-U~\cite{quantenna-coex-sc}. The 3GPP has initiated a feasibility study on the use of 6~GHz bands by 3GPP-based unlicensed RATs~\cite{tr37890}. At the time of writing this paper, however, coexistence mechanisms defined for NR-U are band agnostic, i.e., the same mechanisms are defined for NR-U operations in the 5~GHz and 6~GHz bands. Furthermore, the Technical Report 37.890~\cite{tr37890} does not specify any new mechanisms or features that are tailored for 6~GHz bands. However, with work on future releases of NR-U (such as 3GPP Release 17) already underway, novel mechanisms that leverage the abundant spectrum available in the 6~GHz bands can be designed. Further, the greenfield spectrum creates a unique opportunity for the design of new and fair coexistence mechanisms between NR-U and other unlicensed RATs in the 6~GHz unlicensed spectrum. 

\subsection{Technologies in Adjacent Bands}
\label{subsec:adjacent}

The 6~GHz unlicensed spectrum spans from 5.925--6.425~GHz in Europe and 5.925--7.125~GHz in the US. In both regions, this unlicensed spectrum is adjacent to the 5.9~GHz Intelligent Transportation Systems (ITS) band, which spans from 5.85--5.925~GHz band. The ITS band supports vehicle-to-everything (V2X) communication systems, which cater to many safety and non-safety vehicular communications applications~\cite{carter2005status}. The safety-of-life nature of these applications necessitates the protection of ITS band services from all sources of external interference. In Sec.~\ref{sec:adjacent}, we discuss the impact of unlicensed operations in the 6~GHz bands on the performance of ITS band technologies. In what follows, we briefly introduce the candidate V2X technologies that can operate in the 5.9~GHz ITS band.

\subsubsection{IEEE 802.11p and IEEE 802.11bd}

The first technology to enable direct vehicular communications---Dedicated Short Range Communications (DSRC)---was developed nearly two decades ago~\cite{kenney2011dedicated}. The design objective of DSRC was to support basic safety applications~\cite{carter2005status}, where the DSRC on-board unit module provided alerts to drivers who can then make corrective maneuvers, thereby avoiding potential crashes and accidents. DSRC is based on the IEEE 802.11p standard for its PHY and MAC layers, which in turn is derived from the IEEE 802.11a standard. DSRC has undergone several years of development, planning, and testing through numerous pilot test-beds~\cite{ameixieira2014harbornet, santa2017deployment, chowdhury2018lessons, ng2018besafe}. The performance of 802.11p makes it suitable to support most safety applications envisioned during the conception of DSRC---referred to as day-1 vehicular safety applications. However, for certain advanced applications that simultaneously require high reliability and low latency, 802.11p's performance remains unsatisfactory~\cite{morgan2010notes}. In order to close the gap between application requirements and 802.11p performance, the work on the evolution of 802.11p---IEEE 802.11bd---has now begun~\cite{naik2019ieee, bazzi2019survey}. IEEE 802.11bd will leverage PHY and MAC layer enhancements made by the Wi-Fi community since the design of 802.11p, i.e., features that have been incorporated in 802.11n and 802.11ac.

\subsubsection{Cellular V2X and New Radio V2X}

Cellular V2X (C-V2X) is a RAT for vehicular communications standardized by the 3GPP in its Release 14~\cite{wang2017overview}. The PHY layer of C-V2X is based on LTE. Depending on whether the User Equipments (UEs), i.e., vehicles, are within cellular coverage or outside, the C-V2X sidelink mode 3 or mode 4, respectively, is used for resource allocation~\cite{molina2017lte}. The sidelink interface was designed by the 3GPP for direct device-to-device communication in its Release 12~\cite{lien20163gpp}. In sidelink mode 3, the resource assignment for individual UEs is performed by the cellular base station, i.e., the eNodeB~\cite{bazzi2017performance}. On the other hand, the 3GPP has defined a semi-persistent resource reservation algorithm~\cite{molina2017lte}, using which UEs can reserve resources in a distributed fashion without assistance from the cellular infrastructure. The objective of C-V2X sidelink modes 3 and 4, similar to that of DSRC, is to enable support for day-1 vehicular safety applications. However, similar to DSRC, the Release 14-based C-V2X faces shortcomings in extremely dense vehicular settings~\cite{toghi2018multiple, naik2020c}. Furthermore, the latency guarantees provided by C-V2X are incapable of meeting the requirements of advanced vehicular applications. Thus, work is underway in 3GPP for the development of C-V2X's evolution---NR V2X~\cite{naik2019ieee}.

\subsection{6~GHz channelization}
\label{subsec:channelization}

The list of unlicensed channels in the 6~GHz bands in the US and Europe are shown in Fig.~\ref{fig:6GHz-channels-us} and Fig.~\ref{fig:6GHz-channels-europe}, respectively. The number of 6~GHz channels available in the US and Europe is shown in Table~\ref{table:6GHz-channels}. The relationship between the center frequency of a channel and the corresponding channel number in the 6~GHz band, as adopted by the IEEE 802.11ax specifications~\cite{11axDraft} and the European Communications Committee (ECC) Report 302~\cite{ecc-report-302} is given by Eq.~\eqref{eqn:6GHz-channels}.

\begin{equation}
    f_c (g) = 5940 + (g \times 5) \quad \textrm{MHz},
    \label{eqn:6GHz-channels}
\end{equation}

where $g$ is the channel number and $f_c(g)$ is the center frequency of that channel. The 20~MHz channels shown in Fig.~\ref{fig:6GHz-channels-us} and Fig.~\ref{fig:6GHz-channels-europe} are given by channel numbers $1, 5, 9 \cdots$, while the 40~MHz channels are given by channel numbers $3, 11, 19, \cdots$. Channel numbers for 80, 160 and 320~MHz channels are shown in Fig.~\ref{fig:6GHz-channels-us}.

\begin{table}[htb]
    \centering
    \begin{tabular}{|c|c|c|}
        \hline
        \textbf{Bandwidth} & \textbf{United States} & \textbf{Europe} \\ \hline
        20~MHz & 59 & 24 \\ \hline
        40~MHz & 29 & 12 \\ \hline
        80~MHz & 14 & 6 \\ \hline
        160~MHz & 7 & 3 \\ \hline
        320~MHz & 3 & 1 \\ \hline
    \end{tabular}
    \caption{Number of 6~GHz channels in the US and Europe.}
    \label{table:6GHz-channels}
\end{table}

%% file: opportunities.tex
\section{Benefits of the 6 GHz Spectrum}
\label{sec:opportunities}

\subsection{Throughput Enhancement}
\label{subsec:throughput}

Allowing unlicensed access in the 6~GHz bands opens up as much as 1.2~GHz of additional spectrum in the US and 500~MHz of spectrum in Europe. In the US, this additional spectrum more than doubles the amount of spectrum currently available for unlicensed access across the 2.4~GHz and 5~GHz bands. Naturally, the availability of such a large amount of spectrum for unlicensed access is expected to significantly boost the achievable end-to-end throughput. 

IEEE 802.11be is referred to as Extremely High Throughput, or EHT. In order to live up to this epithet, 802.11be will be required to substantially raise the peak achievable throughput in comparison to the previous 802.11 standards. The theoretical peak throughput of 802.11be is estimated~\cite{11-19-0754} at 207~Gbps! This can be achieved by aggregating nine 160~MHz links using MLA, 16 spatial streams and 4096 QAM. On the other hand, if aggregation is infeasible, 802.11be devices can achieve up to 46~Gbps using a single 320~MHz channel, 4096 QAM, and 16 spatial streams~\cite{11-19-0754}. These peak throughput numbers are substantially higher than those promised by previous Wi-Fi generations. 

\subsubsection{Wider Bandwidths}
IEEE 802.11be is likely to extend the permissible set of channel bandwidth to include 240~MHz and 320~MHz channels~\cite{11-19-1889r2}. The current Wi-Fi 6 (i.e., 802.11ax-certified) and prior generation of Wi-Fi devices allow support for channel bandwidth of 20, 40, 80, or 160~MHz. A 240~MHz channel can be created by combining a 160~MHz channel with another 80~MHz channel. This is denoted as 160+80~MHz. A 240~MHz channel can also be created by combining three 80~MHz channels (indicated as 80+80+80~MHz). Similarly, 320~MHz channels can be created as 80+80+80+80~MHz, 160+80+80~MHz or 160+160~MHz channels. It must be noted that the combined channels need not be contiguous, thereby providing maximum flexibility in creating different Wi-Fi channel configurations.

Even though the creation of 320~MHz channels was theoretically possible in the 5~GHz bands, only a single 320~MHz was available in the 5~GHz bands, thereby severely limiting its use. The 6~GHz unlicensed bands, on the other hand, open up as many as three~160~MHz, two~240~MHz, and one~320~MHz channel(s) in Europe and seven 160~MHz, four~240~MHz and four~320~MHz channels in the US! Thus, even if 802.11be extends support for 240~MHz and 320~MHz channels, such channels can be meaningfully used only in the 6~GHz bands.

\begin{figure}
    \begin{subfigure}{0.5\textwidth}
     \centering
     \includegraphics[trim={5.7cm 1.6cm 12cm 12.5cm},clip,scale=0.75,angle=-90]{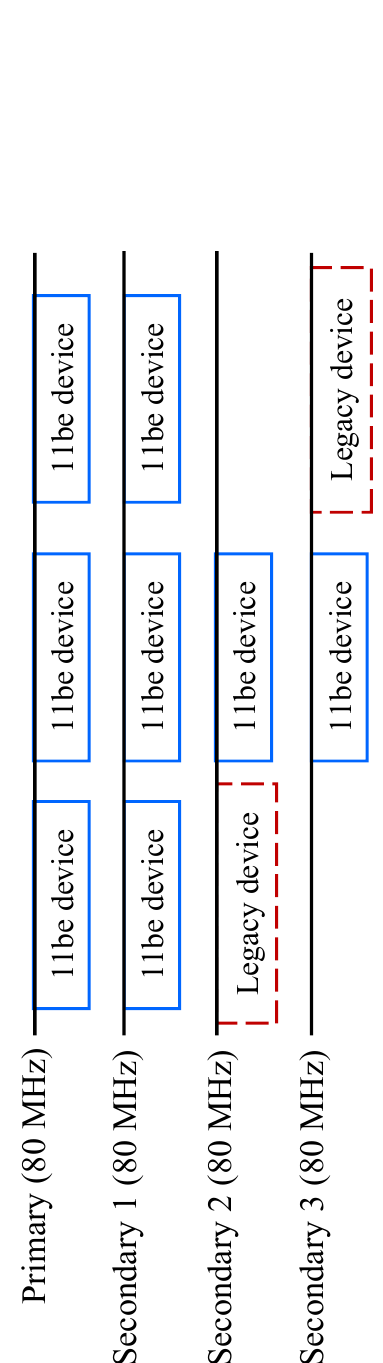}
     \subcaption{Without 240 MHz channelization.}
     \label{fig:without240}
    \end{subfigure}\\
    \begin{subfigure}{0.5\textwidth}
     \centering
     \includegraphics[trim={10.6cm 1.6cm 6.9cm 12.5cm},clip,scale=0.75,angle=-90]{figures/240-MHz.pdf}
     \caption{With 240 MHz channelization.}
     \label{fig:with240}
    \end{subfigure}
    \caption{Benefits of allowing 240~MHz channels in IEEE 802.11be.}
    \label{fig:240-320}
    \vspace{-4mm}
\end{figure}

Note that enabling support for 240~MHz channels deviates from 802.11's previous practice of supporting channel bandwidth in the form $\textrm{W} = 20 \times 2^{i}$, where $\textrm{W}$ is a channel bandwidth supported by a Wi-Fi device and $i = 0, 1, 2, 3, 4$ for $\textrm{W} = 20, 40, 80, 160, 320$~MHz, respectively. This decision has been consciously taken because providing support for both 240~MHz and 320~MHz channels yields higher throughput than only supporting 320~MHz channels~\cite{11-19-0778r0}. This is because allowing 240~MHz channels unlocks additional channel options in cases where certain portions of a 320~MHz channel are busy. This is exemplified in Fig.~\ref{fig:240-320}. As seen in Fig.~\ref{fig:240-320}, if 240~MHz channels are not allowed, if either of secondary channels 1, 2, or 3 are busy, transmissions can occur only on 80 or 160~MHz channels. However, if 240~MHz channels are allowed and if only one of the constituent 80 MHz channel is busy, the remaining 80 MHz channels can be bonded to form a 240~MHz channel, thereby resulting in higher throughput. Note than bonding non-contiguous 80~MHz channels is already permitted in IEEE 802.11ac. 

\begin{figure}[htb]
    \centering
    \includegraphics[trim={5.7cm 1.5cm 7.8cm 10.7cm},clip,scale=0.6,angle=-90]{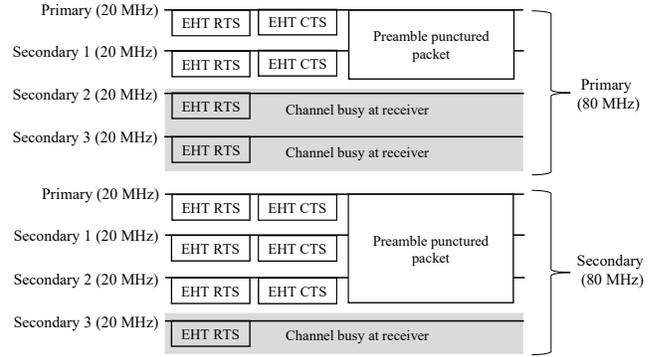}
    \caption{Improving spectral efficiency using preamble puncturing in IEEE 802.11be.}
    \label{fig:mu-rts}
\end{figure}

In addition to allowing 240~MHz channels, in order to increase the spectral efficiency IEEE 802.11be will use preamble puncturing to \textit{puncture} data transmission in all those 20~MHz portions of a large channel that are busy and transmit in the remaining portions of the channel~\cite{11-19-2125r0}. This is supported using EHT Request-to-Send (RTS) and EHT Clear-to-Send (CTS) frames, as shown in Fig.~\ref{fig:mu-rts}. It must be noted that optional support for preamble puncturing has already been enabled in IEEE 802.11ax~\cite{11axDraft}.

\subsubsection{Higher Order MCS \& Preamble Design}
Besides enabling access to larger channel bandwidths, the 6~GHz bands assist in provisioning higher throughput in a few other ways. Firstly, 802.11be shall, for the first time, allow the use of 4096 QAM~\cite{11-19-0637}. Decoding densely packed symbols of 4096 QAM necessitates a very high signal-to-interference-plus-noise ratio (SINR) at the receiver. Such high SINRs are hard to achieve in the 2.4~GHz and 5~GHz bands due to a large number of ad-hoc Wi-Fi deployments and other sources of interference (e.g., microwaves in the 2.4~GHz band and LTE-LAA in the 5~GHz bands). On the other hand, with access to large amounts of spectrum in the 6~GHz bands, the large number of channels therein is likely to enable better frequency planning, which can minimize interference and support the necessary SINR for decoding 4096 QAM symbols. 

Secondly, the 6~GHz bands provide a unique opportunity where 802.11ax and 802.11be devices need not be backward compatible with previous Wi-Fi generations (802.11ac/n/a/g). Consequently, 802.11ax and 802.11be devices, when operating in the 6~GHz bands, can cease to transmit legacy fields in their preambles. These fields include the \textit{Legacy Short Training Field}, \textit{Legacy Long Training Field} and \textit{Legacy Signal}, which are used by all IEEE 802.11 devices for backward compatibility in the 2.4~GHz and 5~GHz bands~\cite{11-19-1556r1}. The air-time overhead reduction resulting from not transmitting these legacy preamble fields might not seem much but is significant because of two primary reasons: (i) Nearly two-thirds of all frames transmitted by Wi-Fi devices are control and management frames whose payload sizes are very small. Since the length of these frames is on the same order as the size of legacy preamble fields, the overhead resulting from transmission of legacy preamble fields becomes significant~\cite{11-19-1556r1}, and (ii) For increased robustness, these preamble fields are transmitted at the lowest MCS. Therefore, the time taken to transmit these fields is considerable. For example, in the amount of time required to transmit the legacy preamble fields, an IEEE 802.11be device could transmit up to 1~Mb of data using MCS-11 and 16 spatial streams over a 320~MHz channel~\cite{11-19-1556r1}!

\subsubsection{Impact of MLA on Throughput}
In addition to the features mentioned above, IEEE 802.11be devices can use MLA to aggregate channels in the same or different bands. As described in Sec.~\ref{subsubsec:wi-fi}, MLA will allow 802.11be devices to transmit packets belonging to the same access category on two channels/links simultaneously (either in a synchronous or asynchronous manner). Note that from a throughput enhancement viewpoint, the larger the channel bandwidth available, the better. Theoretically, if two channels of the same bandwidth are aggregated, the achieved throughput can be doubled. While only 20~MHz and 40~MHz channels are available in the 2.4~GHz bands, the 5~GHz bands have several 80~MHz and 160~MHz channels available. Thus, for boosting the achievable throughput, it is desirable to aggregate two or more wide-bandwidth channels in the 5~GHz and/or 6~GHz bands. 

Consider an example where channel 42 (80~MHz channel in the U-NII-1 band) is aggregated with channel 7 (the first 80~MHz channel in the U-NII-5 band). The significant frequency separation between the two aggregated channels can potentially eliminate the inter-link cross-talk, thereby allowing for the use of asynchronous/independent MLA to aggregate the two channels. Based on traffic conditions, simulation studies have shown that independent/asynchronous MLA used in such settings can achieve up to 2 to 4 times of the single-link throughput~\cite{11-19-0764r1}. On the other hand, channel 155 (80~MHz channel in the U-NII-3 band) can be aggregated with channel 7 (the first 80~MHz channel in the U-NII-5 band). The small frequency separation between the two aggregated channels, in this case, can lead to significant cross-talk between the two channels. Synchronous multi-primary MLA, which is suitable for aggregation in such cases, can boost the throughput by up to 5 times in heavy traffic conditions as reported in references~\cite{11-19-0766r1, 11-19-1291r3}. 

Note that synchronous multi-primary MLA yields higher performance gains compared to asynchronous/independent MLA. The additional gain comes from the \textit{free-riding} of the secondary link, i.e., the secondary link that is aggregated when ED check passes for PIFS interval of time. Since the MLA device gains access to this secondary link without undergoing the entire contention procedure (i.e., IFS + backoff), the device is said to acquire a free-ride on this secondary link. Although this free-riding is advantageous for MLA-capable devices, it leads to unfairness towards single link and legacy 802.11 devices operating on the secondary link. We discuss this issue further in Sec.~\ref{sec:challenges}.


\subsection{Latency Enhancement}
\label{subsec:latency}

\begin{figure}[htb]
    \centering
    \includegraphics[trim={6.85cm 9cm 7cm 7cm},clip,scale=0.73,angle=-90]{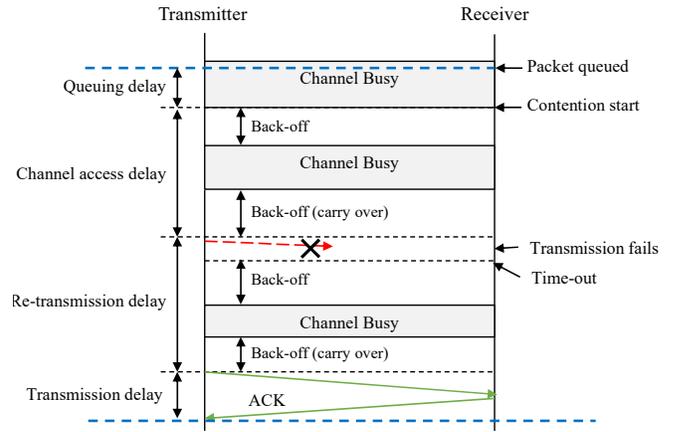}
    \caption{Components of delay in packet transmission in an LBT-based system.}
    \label{fig:latency-components}
\end{figure}

In LBT-based unlicensed RATs such as Wi-Fi and NR-U, the end-to-end latency of packet transmissions is composed of four major components. These are the delays encountered in packet queuing, channel access, transmission, and re-transmission\footnote{These are the four major components in an LBT-based system that can take values of up to tens or hundreds of milliseconds. In practice, delays are also encountered due to propagation over the medium and processing. However, these delays can be ignored because they are much smaller than the ones considered above.}. These latency components are illustrated in Fig.~\ref{fig:latency-components}~\cite{11-19-1938r2, 11-19-1942r7}.

A packet experiences queuing delay if it is not at the head of the queue when generated. All preceding packets must be transmitted successfully before such a packet can be placed at the queue-head. This constitutes the packet queuing delay. Thereafter, the packet must wait in the queue until the contending device wins access to the medium. This delay constitutes channel access latency. Upon winning access, if the transmitted packet is successfully received at the receiver, the only further delay experienced in its transmission is the transmission delay (i.e., there is no further latency addition due to re-transmission(s)), which depends on the channel capacity. However, if the packet transmission fails either due to collision(s) or poor channel conditions, the packet must be re-transmitted. Each re-transmission further introduces channel access and transmission delays. In order to minimize the cumulative end-to-end latency, each of the four constituent delays must, therefore, be minimized. 

In lightly loaded environments, where the number of contending users is small or if the traffic generated by these contending users is sporadic, packet collisions are infrequent. In such scenarios, delays due to queuing and re-transmission(s) are negligible, and the end-to-end latency of packet transmissions is dominated by channel access and transmission delays. Although transmission delay can be minimized with faster transmission rates (i.e., higher-order MCS), increasing the PHY rate has negligible impact on the channel access latency. On the other hand, the channel access latency, especially in the worst case, is dominated by the LBT-based contention process used by Wi-Fi and NR-U devices. It has been identified that the worst-case channel access latency is the most significant bottleneck encountered in enabling real-time applications~\cite{11-19-0065r6}.

Since Wi-Fi 6 and future generations of Wi-Fi will support the schedule-based MU-OFDMA mode for uplink transmissions, efficient scheduling at the AP can be used to reduce packet latencies. In addition to increasing the offered throughput~\cite{bhattarai2019uplink}, the authors in~\cite{avdotin2019enabling} demonstrate that simple modifications to the default channel access rules in MU-OFDMA can go a long way in improving the latency, thereby enabling real-time applications.

\subsubsection{Impact of MLA on Latency}
MLA in IEEE 802.11be has the potential to significantly bring down the worst-case channel access latency. This has been demonstrated in reference~\cite{11-19-1081r1} for independent/asynchronous MLA. Even with a modest number of full-buffer contenders, the worst case (95 percentile) latency can be of the order of hundreds of milliseconds for single link 802.11 devices. This is because repeated transmissions from full-buffer contenders can potentially result in repeated collisions, which can quickly push the CW values of the contending devices to very large values, thereby increasing the channel access latency. However, when 802.11be devices are allowed to aggregate two links, there is an order of magnitude reduction in the 95 percentile latency. Similar results have been demonstrated in reference~\cite{11-19-0979r2}, where it can be seen that asynchronous/independent MLA can reduce the latency to up to 25~\% of the single link case. 

\begin{figure}
    \begin{subfigure}{0.5\textwidth}
     \centering
     \includegraphics[trim={4.75cm 6.2cm 13.6cm 10.7cm},clip,scale=0.72,angle=-90]{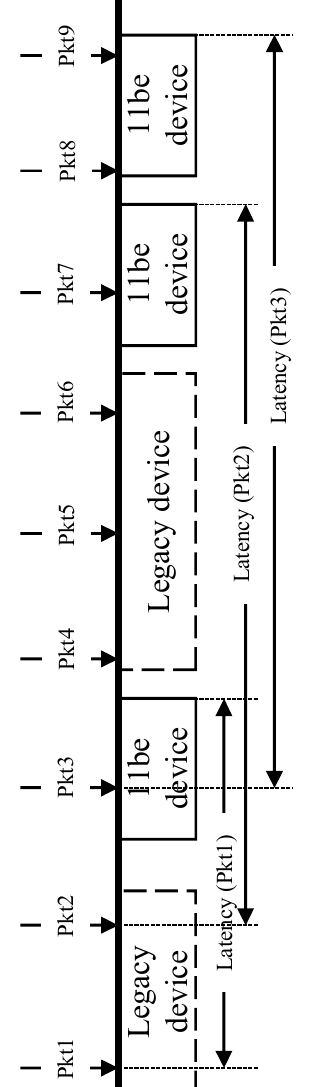}
     \subcaption{Latency in single link 802.11 networks.}
     \label{fig:latency-single-link}
    \end{subfigure}\\
    \begin{subfigure}{0.5\textwidth}
     \centering
     \includegraphics[trim={9cm 6.2cm 7.5cm 10.7cm},clip,scale=0.72,angle=-90]{figures/MLA-benefits.pdf}
     \caption{Latency in MLA-enabled 802.11 networks.}
     \label{fig:latency-mla}
    \end{subfigure}
    \caption{Latency enhancement with asynchronous MLA.}
    \label{fig:latency-enhancement}
    \vspace{-4mm}
\end{figure}

With MLA, packets are flushed from the queue as soon as one of the links becomes idle. Thus, unlike single link transmissions, where a contending device must wait for the sole link to be idle, if any of the links on which the MLA device can transmit becomes idle, the pending packet can be transmitted. This is shown in Fig.~\ref{fig:latency-enhancement}. Observe that even in this dummy example with just two links available for contention, the latency of packets is substantially reduced as packets are flushed out as soon as one of the links becomes idle. It is easy to infer that larger the number of links that can be aggregated by an MLA device, lower is the worst-case latency~\cite{11-19-0760r1}. It is worthwhile to note from~\cite{11-19-1081r1} that link parameters such as MCS or bandwidth have far less of an impact in comparison to the availability of an additional link on which the MLA device can contend. Thus, unlike throughput enhancement, the latency performance substantially improves even if an 80/160~MHz channel is aggregated with a 20~MHz channel. 

It is, therefore, clear that when using a feature like MLA, the single most important factor that is responsible for reducing the worst-case latency is the availability of additional links/channels on which an MLA-capable device can contend. With the opening up of 6~GHz bands, the number of available channels increases substantially (for both synchronous and asynchronous MLA), thereby significantly contributing toward enabling support for latency-sensitive real-time applications. The reduced latency resulting from the use of MLA has an impact on not only the first transmission of the packet but also its subsequent re-transmission(s) if the first transmission fails. 

\subsubsection{Re-transmissions}
An alternate mechanism to reduce the latency induced by packet re-transmissions is to transmit latency-sensitive packets multiple times within the latency budget without waiting for feedback (i.e., ACK) from the receiver~\cite{11-19-1622r0, 11-19-1884r1}. This approach has an added benefit in that the reliability of packet transmissions increases considerably; Even if one (or more) packet transmission(s) fail, the probability that all transmissions of a packet fail becomes vanishingly small. Thus, using this re-transmission mechanism, packets are reliably delivered in under their latency budget. Note that this approach is similar to the re-transmission approach being considered in NR V2X, where low latency and high reliability is a critical requirement~\cite{naik2019ieee}. The drawback of this approach is that multiple transmissions of each packet lead to significant capacity requirements, which is where the availability of 6~GHz bands becomes of utmost importance.

\subsubsection{Greater sub-carrier spacing in NR-U}
NR-U devices can leverage greater sub-carrier spacing to reduce packet latencies. As the sub-carrier spacing increases, the symbol duration reduces, thereby enabling NR-U devices to transfer the same amount of data in a smaller duration~\cite{ahmadi}. Further, greater sub-carrier spacing combined with mini-slot scheduling can reduce the amount of time spent in transmitting reservation signals, which not only increases the NR-U air-time efficiency but also allows devices to transmit packets as soon as possible~\cite{ahmadi}. Since the 6~GHz bands are greenfield spectrum, the complexities associated with coexistence across different numerologies can be eliminated if wider sub-carrier spacings are adopted right at the onset of NR-U operations in the 6~GHz bands. However, while doing so, care must be taken that the performance degradation due to increased multi-path fading at greater sub-carrier spacing, especially in outdoor environments, does not counteract the benefits of improved latency.

%% file: co-channel-incumbent.tex
\section{Coexistence with Incumbent Technologies}
\label{sec:incumbent}

\subsection{Coexistence in the US}

\begin{table*}[htb]
    \centering
    \begin{tabular}{|c|c|c|c|}
        \hline
        \textbf{Device class} &  \textbf{Bands} & \textbf{Maximum EIRP} & \textbf{Maximum EIRP PSD}\\ \hline
        Standard power AP & \multirow{2}{*}{U-NII-5/U-NII-7} & 36~dBm & 23~dBm/MHz \\
        Clients connected to standard power AP & & 30~dBm & 17~dBm/MHz \\ \hline
        LPI AP & \multirow{2}{*}{U-NII-5 through U-NII-8} & 30~dBm & 5~dBm/MHz \\ 
        Clients connected to LPI AP & & 24~dBm & -1~dBm/MHz \\ \hline
        VLP devices$^*$ & U-NII-5 through U-NII-8 & 4 to 14~dBm & -18 to -8~dBm/MHz \\ \hline
    \end{tabular}
    \caption{Unlicensed devices allowed to operate in the 6~GHz bands in the US. $^*$ indicates device classes considered in the FCC FNPRM~\cite{fcc-ro}. Note that throughout the 6~GHz bands, deployment of unlicensed devices in moving vehicles, trains and unmanned aerial vehicles is prohibited.}
    \label{table:unlicensed-classes}
\end{table*}

\subsubsection{U-NII-5 and U-NII-7 bands}
\label{subsec:5-7}

The FCC 6~GHz R\&O~\cite{fcc-ro} allows standard power APs\footnote{Here, the term AP, as used in the FCC NPRM and R\&O, refers to both Wi-Fi AP and NR-U gNB.} (23~dBm/MHz) to operate in the U-NII-5 and U-NII-7 bands under the constraints that the APs must not operate within the \textit{exclusion zone} of incumbent receivers. The exclusion zone for incumbent receivers is computed as the region around the receiver where unlicensed operations can result in interference-to-noise-power ratio, I/N = -6~dB, or greater~\cite{fcc-ro}. Exclusion zones for incumbent receivers operating in U-NII-5 and U-NII-7 bands will be computed by a centrally-implemented AFC system. These computations will take into account the antenna pattern of the incumbent receiver, antenna heights of the incumbent and unlicensed users, the Effective Isotropic Radiated Power (EIRP) of the unlicensed devices, and appropriate propagation models. 

\begin{figure}[htb]
    \centering
    \includegraphics[trim={5cm 9.3cm 7.95cm 9.4cm},clip,scale=0.7,angle=-90]{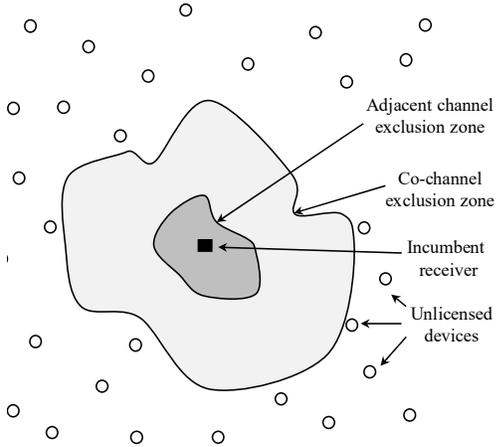}
    \caption{Co-channel and adjacent channel exclusion zones.}
    \label{fig:exclusion-zones}
\end{figure}

To mitigate adjacent channel interference resulting from unlicensed transmitters at a given incumbent receiver, the AFC database will compute not only the co-channel (i.e., the channel on which the incumbent receiver operates) exclusion zone but also the adjacent-channel exclusion zone. The concept of co-channel and adjacent channel exclusion zones is illustrated in Fig.~\ref{fig:exclusion-zones}. Suppose the incumbent user operates on channel $c$. Unlicensed devices are then prohibited from operating on channel $c$ (or channels that overlap partially with channel $c$) inside the co-channel exclusion zone. Further, unlicensed devices are prohibited from operating on any channel that is immediately adjacent to channel $c$ within the adjacent channel exclusion zone. Note that because wireless devices radiate less power into adjacent channels than in the intended channel (Fig.~\ref{fig:spectral-mask}), the exclusion zone for adjacent channels is smaller than the co-channel exclusion zones. 

A standard power unlicensed AP, before beginning operations in the U-NII-5 and U-NII-7 bands, must contact the AFC system database and obtain a list of permissible frequencies for operations. To facilitate the computation of this list, the AP must provide its location and antenna height. This requirement necessitates either the possession of geo-location capabilities (such as Global Positioning System, GPS) at APs operating in the U-NII-5 and U-NII-7 bands or access to an external geo-location source. In addition to its location, the AP must report the uncertainty in its location (i.e., accuracy in meters) to the AFC system. Further, if the AP's geo-location is obtained through an external source, it must report its distance from this external source as a part of its location uncertainty. Because incumbent users in these bands are static and new users become operational infrequently, the FCC R\&O requires unlicensed APs to query the database and obtain a list of permissible frequencies once every 24 hours.

The AFC system database will provide a set of transmission powers that the standard power AP can select. For each transmission power in this set, a corresponding list of permissible frequencies will be provided. The FCC R\&O states that the AFC system must compute, at a minimum, the list of frequencies for transmission power in increments of 3~dB from 21~dBm to 36~dBm~\cite{fcc-ro}. The unlicensed AP can then operate on one of these permitted frequencies. Unlicensed clients (STAs for Wi-Fi and UEs for NR-U) can operate in the 6~GHz bands only when associated with an AP. Note that both indoor and outdoor operations of standard power APs are permitted as long as the AFC system database indicates that the given frequency is available at the AP's location. 

In addition to standard power APs, the FCC R\&O allows low-power APs (5~dBm/MHz) and clients (-1~dBm/MHz) to operate throughout the 6~GHz bands without contacting the AFC system as long as these devices are confined to indoor environments~\cite{fcc-ro}. This implies that as long as such APs are indoors and operate withing the permissible power levels, they can operate inside the exclusion zones of incumbent receivers. To ensure that such APs---referred to as Low Power Indoor (LPI) APs---operate only in indoor settings, the FCC has placed the following three constraints: (i) LPI APs \textit{cannot} be weather resistant, (ii) LPI APs must be equipped with integrated antennas and the capability of connecting such APs with external antennas is prohibited, and (iii) LPI APs cannot operate on battery power. 

The FCC Further Notice of Proposed Rulemaking (FNPRM)~\cite{fcc-ro} additionally seeks comments on allowing unlicensed devices operating at extremely low powers throughout the 6~GHz bands. These devices---referred to as very low power (VLP) devices---can operate within the exclusion zones of incumbent users and in both indoor and outdoor environments. Such VLP devices are critical in enabling personal area network applications such as wireless AR and VR through wearable devices and in-car connectivity~\cite{vlp-use-cases}. Although the power spectral density (PSD) for VLP devices is yet to be determined, the FCC FNPRM proposes allowing VLP devices radiating under 4--14~dBm EIRP to operate in all four 6~GHz sub-bands. Note that due to building entry losses, the median value of which is 20.25~dB~\cite{fcc-ro}, LPI devices are permitted to operate within the exclusion zones of incumbent users. The choice of EIRP for VLP devices (4--14~dBm) is motivated by the fact that VLP devices radiating in outdoor environments and within the exclusion zones of incumbent receivers must at least compensate for the building entry losses faced by signals from LPI devices. 

Table~\ref{table:unlicensed-classes} shows the different categories of unlicensed devices permitted to operate in the 6~GHz bands in the US~\cite{fcc-ro}. Note that for standard power devices, considering the PSD limits per MHz of spectrum, the maximum EIRP is achieved for 20~MHz channels (23~dBm/MHz + 10 $\times$ log$_{10}$ (20) = 36~dBm). This implies that if the bandwidth is increased beyond 20~MHz (say to 40~MHz), the power per sub-carrier is reduced by the corresponding factor (halved for 40 MHz). For LPI and VLP devices, the maximum EIRP is achieved for 320~MHz (5~dBm/MHz + 10 $\times$ log$_{10}$ (320) = 30~dBm) and 160~MHz (-8~dBm/MHz + 10 $\times$ log$_{10}$ (160) = 14~dBm), respectively. Further, note that the EIRP limit for clients is 6~dB is lower than those for the corresponding APs. This is due to the following two reasons: (i) Restricting the transmission power of client devices reduces their transmission range. This ensures that client devices will operate close to their respective APs. This is especially important in the U-NII-5 and U-NII-7 bands, where the permissible frequencies of operation are closely tied to the location of devices. (ii) Client devices are typically battery-operated and hence, power-limited. This reduces the incentive of client devices to transmit at higher power levels. 

While AFC-like database systems have previously been used for enabling coexistence between incumbent and unlicensed devices~\cite{grissa2019trustsas}, studies on the impact of Wi-Fi devices on the performance of fixed services operating in the 6~GHz bands have so far been inconclusive. The Fixed Wireless Communications Coalition estimates that if Wi-Fi devices are permitted to operate within the incumbent users' exclusion zone, fixed service links can fail entirely with 1.6\% probability. At the same time, bit errors can occur with 7.1\% probability~\cite{ex-parte-fwcc-indoor}. The coalition further argues that because fixed service links have very stringent reliability requirements, studies on the impact of unlicensed devices on incumbent services must perform worst-case interference analysis. For example, instead of assuming \textit{typical} terrain and clutter conditions, simulation studies must use free space path loss models to compute the interference power at incumbent receivers~\cite{ex-parte-fwcc, ex-parte-ctia-2}. Additionally, incumbent operators have argued for mandatory AFC system requirements for not only standard power devices (as indicated in the R\&O) but also LPI devices~\cite{ex-parte-ctia-2, ex-parte-fwcc-indoor, ex-parte-ctia}. 

Notwithstanding the above arguments, several studies from proponents of unlicensed services have dismissed incumbent users' concerns~\cite{ex-parte-vlp, ex-parte-vlp-2,ex-parte-vlp-3, ex-parte-6usc, ex-parte-benchtop, ex-parte-vlp-4}. Specifically, reference~\cite{ex-parte-6usc} notes that for standard power unlicensed services to interfere with incumbent users, a series of unlikely events must co-occur, thereby indicating that the probability of harmful interference from unlicensed users is extremely small. Results from bench-top testing performed in reference~\cite{ex-parte-benchtop} indicate that when the interference protection criterion is fixed to I/N=-6~dB, the presence of Wi-Fi devices does not impact the performance at fixed service receivers. Furthermore, using simulation studies reference~\cite{ex-parte-vlp-4} shows that harmful interference resulting from LPI devices at fixed service links is extremely unlikely. Studies performed in~\cite{ex-parte-hp-broadcom} for LPI devices located in high rises of New York City and Washington DC yield similar conclusions. Before the FCC R\&O was released, results from these studies were taken into consideration. Since the FCC R\&O allows for standard power APs to operate in the U-NII-5 and U-NII-7 bands and LPI devices to operate throughout the 6~GHz bands, it can be inferred that the FCC concludes that studies from proponents of unlicensed services are more practical and representative of real-world coexistence environments. 

\subsubsection{U-NII-6 and U-NII-8 bands}
\label{subsec:6-8}

As noted in Sec.~\ref{subsec:incumbent}, incumbent users of the U-NII-6 and U-NII-8 bands are such that their exact locations are hard to determine at a given time. This renders an AFC-like database system inefficient in protecting such incumbent users from interference resulting from unlicensed operations. Consequently, in these bands, the FCC R\&O allows only LPI operations and seeks comments in its FNPRM on whether VLP devices can be permitted in outdoor environments. The constraints set by the FCC for ensuring that unlicensed devices operating in the U-NII-6 and U-NII-8 bands are restricted to indoor environments are the same as those outlined in Sec.~\ref{subsec:5-7}. 

By restricting unlicensed devices in indoor environments, it is expected that the additional propagation losses, e.g., building entry loss, which is typically on the order of 20~dB~\cite{fcc-ro}, will considerably reduce the strength of unlicensed interference at incumbent receivers. Nevertheless, just like in the U-NII-5 and U-NII-7 bands, proponents of unlicensed services have faced resistance in the U-NII-6 and U-NII-8 bands from its incumbent and existing users. For example, using simulation studies reference~\cite{ex-parte-nab} shows that roughly two-thirds of television pick-up cameras in indoor environments and half of the cameras in outdoor settings can suffer from harmful interference in the presence of unlicensed users. Further, reference~\cite{ex-parte-alion} analyzes Wi-Fi-induced interference to electronic newsgathering systems. The analysis, which is performed for three representative locations, shows that the interference increases with an increase in Wi-Fi duty cycle and antenna height of the newsgathering stations. 

Similarly, existing users of these bands that rely on UWB have expressed concerns regarding unlicensed device-induced harmful interference, thereby significantly compromising their performance~\cite{ex-parte-zebra}. Although UWB system users are not entitled to protection from external sources of interference, if new unlicensed operations in the U-NII-6 and U-NII-8 bands deteriorate their system performance, the resulting impact on applications that are already deployed will be significant.

In comparison to the U-NII-5/7 bands, studies defending unlicensed operations in the U-NII-6/8 bands are fewer in number. Monte Carlo simulation studies reported in~\cite{ex-parte-cableLabs} have shown that restriction of Wi-Fi devices to indoor environments is effective in mitigating interference to Broadcast Auxiliary Services operating in the U-NII-6 and U-NII-8 bands. On the other hand, simulations results reported in~\cite{ex-parte-rkf} acknowledge that unlicensed devices can indeed negatively affect the performance of nomadic incumbent users operating in these bands. However, the fraction of incumbent users that suffer from complete link failure and bit errors are reported to be as low as 0.2\% and 1.1\%, respectively.

\subsection{Coexistence in Europe}

Studies on coexistence between Wi-Fi-like unlicensed users and the existing users of the 5.925--6.425~GHz bands (including adjacent bands) in Europe are reported in the ECC Report 302~\cite{ecc-report-302}. These current users include fixed services, fixed satellite services, ITS services, CBTC, radio astronomy, and UWB systems. Results from minimum coupling loss and Monte Carlo analyses carried out in the report conclude that unlicensed devices, as long as they operate indoors (EIRP 23--24~dBm), present a minimal risk of causing interference at fixed service receivers. Thus, the study concluded that spectrum sharing between unlicensed RATs and fixed services is feasible. 

Further, assuming unlicensed devices' deployment models for the year 2025, the aggregate interference from unlicensed devices at satellite receivers in space was estimated. Considering these future unlicensed deployments, the study concludes that as long as \textit{no more than 5\%} of unlicensed devices operate outdoors, fixed satellite services and unlicensed devices can feasibly share the spectrum\footnote{Note that in the US, the FCC does not take the aggregate interference resulting from outdoor unlicensed devices into consideration. The two primary reasons for this, as stated in the FCC R\&O~\cite{fcc-ro}, are: (i) unlicensed devices have little incentive to radiate upward (in the direction of the satellites), and (ii) computation of aggregate power at the incumbent satellite receivers will complicate the AFC system design.}. In terms of coexistence with incumbent systems operating in adjacent (i.e., ITS) or partially overlapping bands (i.e., CBTC), the report concludes that unlicensed devices can share the spectrum with added constraints. These constraints include tighter spectral masks and restricted operations in the lowermost 6~GHz channels (i.e., channels 1, 3, 7, etc.). 

\begin{table*}[htb]
    \centering
    \begin{tabular}{|p{1.7cm}|p{2.3cm}|p{2.8cm}|p{7.4cm}|}
        \hline
        \multirow{2}{*}{\textbf{Device class}} & \multirow{2}{*}{\textbf{Constraint}} & \multicolumn{2}{|c|}{\textbf{Maximum unlicensed EIRP}} \\ \cline{3-4}
        & & \textbf{Coexisting Incumbent} & \textbf{Permitted EIRP} \\ \hline
        \multirow{4}{*}{LPI} & \multirow{4}{*}{Indoor} & Fixed services & 23--24~dBm \\ \cline{3-4}
        & & Fixed satellite services & 23--24~dBm\\ \cline{3-4}
        & & CBTC & 21.5~dBm/20~MHz (in-band) and -29.5~dBm/5~MHz (out-of-band)\\ \cline{3-4}
        & & ITS & -69 to -36~dBm/MHz (out-of-band) \\ \hline
        VLP & Indoor and outdoor & \multicolumn{2}{l|}{Under consideration} \\ \hline
        Standard & To be considered & \multicolumn{2}{l|}{To be considered, approximately 30~dBm}\\ 
        power & (AFC-like database) & \multicolumn{2}{l|}{}\\ \hline
    \end{tabular}
    \caption{Unlicensed device classes feasible in the 6~GHz bands in Europe~\cite{cept-report-73}. For LPI operations, the maximum EIRP is evaluated on a case-by-case basis for each incumbent system. The final rules will likely be the intersection of all criteria.}
    \label{table:europe}
\end{table*}

Based on the findings presented in~\cite{ecc-report-302}, the European Conference of Postal and Telecommunication (CEPT) approved the CEPT report 73~\cite{cept-report-73}. The recommendations made by this report---submitted in response to the ECC mandate~\cite{ecc-first}---for unlicensed operations in the 6~GHz bands are shown in Table~\ref{table:europe}. Observe that because the nature of existing users operating in the 6~GHz bands in Europe and the US is similar, the proposed permitted device classes, constraints, and solutions for coexistence between the unlicensed RATs and incumbents are also similar. It must be noted that studies conducted in the ECC Report 302~\cite{ecc-report-302} evaluate the technical conditions on unlicensed device operations in the 6~GHz bands. These studies will serve as guidelines for the final technical rules, which are likely to be finalized by July 2020~\cite{ecc-first}. Furthermore, although current studies do not focus on mitigation strategies, if found feasible, outdoor unlicensed operations could be permitted in the future under a geo-location database system constraint (like the AFC system in the US).

%% file: co-channel-unlicensed.tex
\section{Coexistence among Unlicensed Technologies}
\label{sec:unlicensed}
During the design phase of IEEE 802.11ax, the 6~GHz bands were not under active consideration for unlicensed operations. Thus, IEEE 802.11ax was conceived to operate only in the sub-6~GHz bands~\cite{80211ax-PAR}, i.e., the 2.4 and 5~GHz bands and other unlicensed bands, wherever available. However, the latest draft amendment of IEEE 802.11ax~\cite{11axDraft} contains explicit provisions for 802.11ax to operate in unlicensed portions of the 6~GHz bands. Thus, IEEE 802.11ax will be the first IEEE standard to allow unlicensed Wi-Fi operations in the 6~GHz bands. Previous generations of Wi-Fi devices, i.e., IEEE 802.11a/b/g/n/ac, will continue to operate in the 2.4~GHz and/or 5~GHz unlicensed bands. On the other hand, even though the 3GPP Release 16 will continue to provide support for LTE-LAA, the sole 3GPP-based RAT that will operate in the 6~GHz bands is NR-U~\cite{quantenna-coex-sc}. 

Several aspects related to fairness in the coexistence of Wi-Fi and NR-U in the 6~GHz bands are under study in 3GPP, the 802.11 Coexistence SC, and ETSI BRAN. These include the MAC protocols, channel contention parameters, duration of unlicensed transmissions, detection mechanisms, the contention window reset procedure in NR-U, etc~\cite{11-19-1145-02}. However, it is argued that before designing coexistence mechanisms for unlicensed technologies in the 6~GHz bands, the fairness criteria must be revisited~\cite{cableLabs-coex-sc}. The fairness criteria set forth by the 3GPP during the design of LTE-LAA was addressed in~\cite{tr36889} as ``LTE-LAA design should target fair coexistence with existing Wi-Fi networks to not impact Wi-Fi services more than an additional Wi-Fi network on the same carrier, with respect to throughput and latency''. This fairness criterion is heavily biased towards protecting the performance of Wi-Fi networks, while the performance optimization of LTE-LAA networks is not taken into consideration. For example, an LTE-LAA network offering zero throughput would satisfy the above criterion~\cite{11-19-1083r1}. 

\subsection{Channel Access Protocol}
\label{subsec:channel-access-protocol}

When operating in the shared spectrum, the performance of each RAT depends on the MAC protocol used by all coexisting RATs along with their respective MAC protocol parameters. Coexistence between IEEE 802.11ac and LTE-LAA in the 5~GHz bands has been extensively studied at the time of LTE-LAA standardization~\cite{chen2016coexistence, naik2018coexistence}. Since IEEE 802.11ax/802.11be and NR-U are successors to IEEE 802.11ac and LTE-LAA, respectively, insights drawn from the 5~GHz coexistence studies can be leveraged to orchestrate efficient coexistence between Wi-Fi and NR-U (and potentially other future RATs) in the 6~GHz spectrum. 

The design of LTE-LAA emphasized ensuring that the introduction of an LTE-LAA network in the vicinity of existing Wi-Fi networks must not affect them any worse than another Wi-Fi network~\cite{mehrnoush2018analytical}. The most convenient manner in which this objective could be achieved was to design the MAC protocol of LTE-LAA similar to that of Wi-Fi. With this in consideration, LBT---which is fundamentally identical to CSMA/CA used by IEEE 802.11 devices---was chosen as the underlying MAC protocol for LTE-LAA. As the greenfield 6~GHz bands provide an opportunity to investigate and design coexistence mechanisms from the ground up, coexistence between unlicensed RATs in the 6~GHz bands need not consider the CSMA/CA-like LBT mechanisms as the baseline. Although such \textit{modified} LBT-based mechanisms have been proposed in the literature (e.g.,~\cite{song2019cooperative}), at the time of writing this paper the LBT mechanism used in Wi-Fi and LTE-LAA continues to be the baseline for 6~GHz operations. In what follows, we describe key considerations that will govern the performance of all unlicensed RATs operating and coexisting in the 6~GHz bands. 

\subsubsection{LBE v/s FBE}
\label{subsubsec:lbe-fbe}

For more than two decades, Wi-Fi---the predominant WLAN technology---has used the Load Based Equipment (LBE) variant of LBT, which allows transmitters to contend for the medium as soon as the channel becomes idle. However, there exists another option of LBT---Frame Based Equipment (FBE), which is argued to yield superior performance---that allows unlicensed devices to contend for the channel beginning only at synchronized frame boundaries~\cite{wang2017survey}. FBE is especially beneficial for NR-U networks because the frame intervals can be defined to be the same as the NR licensed carrier's slot boundaries. However, the NR-U work item~\cite{RP-182878} states that NR-U systems can use FBE LBT only in environments where the absence of Wi-Fi networks is guaranteed. 

\begin{figure}[htb]
    \centering
    \begin{subfigure}{0.5\textwidth}
    \centering
     \includegraphics[trim={4cm 3cm 11.7cm 2cm},clip,scale=0.47,angle=-90]{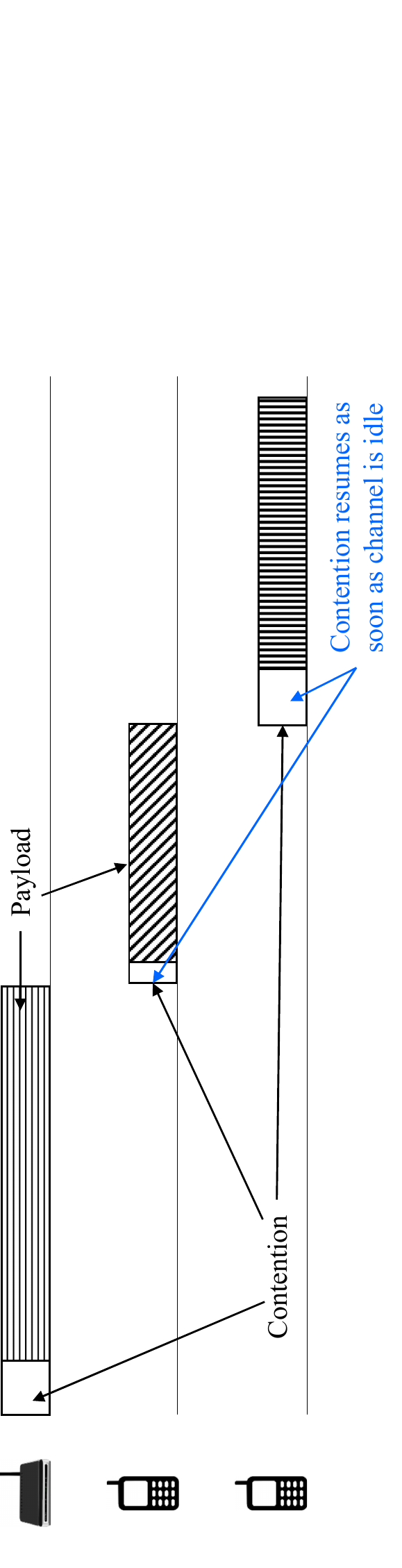}
     \subcaption{Load Based Equipment. Devices contend for the channel as soon as the channel becomes idle.}
     \label{fig:lbe}
    \end{subfigure} \\
    \begin{subfigure}{0.5\textwidth}
    \centering
     \includegraphics[trim={9.7cm 3cm 4cm 3cm},clip,scale=0.47,angle=-90]{figures/lbe-fbe.pdf}
     \subcaption{Frame Based Equipment. Devices contend for the channel only at the beginning of a new frame.}
     \label{fig:fbe}
    \end{subfigure}
    \caption{LBE and FBE flavors of LBT.}
    \label{fig:lbe-fbe}
\end{figure}

Fig.~\ref{fig:lbe-fbe} shows the conceptual differences between LBE and FBE LBT as described in~\cite{etsi-19-104018}. As shown in Fig.~\ref{fig:lbe}, LBE LBT devices can contend for the channel as soon as the previous transmission ends and the channel becomes idle. However, Fig.~\ref{fig:fbe} shows that FBE LBT devices must wait until the next frame interval for contention (and hence, transmission) even if a given transmission does not last until the end of a frame boundary. This implies that FBE LBT devices achieve maximum MAC efficiency when the transmissions last until the end of each frame boundary. However, this may not necessarily be the case, depending on the size of the payload at the transmitter queue. Furthermore, it is important to note that while transmission duration in LBE LBT is limited only by regional regulations, it is the frame interval that limits the transmission duration in FBE LBT, i.e., each transmission must cease by the end of the current frame boundary.

FBE operations for the 5~GHz bands in Europe are defined by ETSI regulations in~\cite{etsi-5GHz}. Although Fig.~\ref{fig:fbe} shows that the contention occurs at the beginning of every frame interval, ETSI regulations specify an alternate version of FBE, where the contending devices perform fixed-duration CCA (category 1 or 2 LBT) towards the end of each frame interval~\cite{R1-1911822, R1-1912088}. In this case, transmitting devices must restrict their maximum channel occupancy time (to 95\% of the frame interval as per~\cite{etsi-5GHz}) so that the end of each frame is idle. The duration of the frame interval can be between 1 and 10~msec. However, once declared, ETSI regulations permit the frame interval to be changed at most every 200~msec~\cite{etsi-5GHz}. Lack of randomization in this version of FBE LBT, where all contending devices perform fixed-duration CCA at the end of each frame interval, makes it susceptible to persistent collisions and is suitable only if interference coordination schemes are used~\cite{R1-1912088}. 

It is well known that if LBE and FBE LBT devices coexist and operate on the same channel, FBE devices get a smaller share of the channel. This is because FBE devices must wait until the next frame boundary, while LBE devices can gain access to the channel immediately after an ongoing transmission ends~\cite{R1-1912088}. Thus, to ensure that LTE-LAA devices are not starved of channel access in the presence of Wi-Fi users, 3GPP adopted the LBE flavor of LBT in LTE-LAA. The NR-U work item, on the other hand, contains explicit provisions to define mechanisms for channel access protocols for FBE LBT, at least in environments where the absence of Wi-Fi networks can be guaranteed~\cite{RP-182878,ahmadi}. It has been proposed that FBE operations in NR-U must be restricted to gNB-initiated channel occupancy time. This would imply that NR-U UEs cannot initiate uplink transmissions using FBE LBT and can transmit in the uplink only when the corresponding gNB wins access to the medium~\cite{R1-1912197, R1-1912257}.

Despite the challenges associated with FBE LBT, it is argued that synchronized access in FBE can yield better performance in comparison to LBE LBT systems~\cite{R1-1912088}. If all unlicensed devices in the 6~GHz bands were to operate using FBE LBT, the spectral efficiency of unlicensed RATs could be increased. With this in consideration, reference~\cite{etsi-19-104018} proposes to use FBE LBT across all unlicensed RATs for synchronized access in the 6~GHz bands. Based on the deployment scenario, it is shown that FBE LBT can improve the user-perceived throughput by 10-50\%. FBE LBT yields performance gains due to the reduction of the hidden nodes problem~\cite{ex-parte-qualcomm}. Since all potential transmitters sense the channel at the same time, i.e., at the beginning of the synchronized frame intervals, the fraction of time during which collisions can occur is reduced and is limited to the start of each frame interval. Furthermore, when transmissions on the channel are synchronized, coordinated interference mitigation and suppression schemes can be leveraged so that multiple devices belonging to the same operator can transmit simultaneously, thereby increasing the spectral efficiency~\cite{R1-1912012}. A critical factor against the use of FBE LBT, however, is that FBE systems require a global clock in order to have consistent frame boundaries across all unlicensed devices. Thus, even though FBE LBT is promising in terms of improved system performance, it is necessary to weigh this improvement against the added system requirement. 

\subsubsection{Contention Parameters}
\label{subsubsec:contention-parameters}

LBE LBT and CSMA/CA, both require a device (with pending packets to transmit) to sense the channel and contend for it before initiating its transmission. If the channel is occupied, i.e., if there is an ongoing transmission, the contending device must wait until the channel is idle. Once the channel is idle, devices enter the well known \textit{exponential back-off} phase, whereby the CSMA/CA or LBE LBT protocol is used (by Wi-Fi and NR-U devices, respectively) to contend for the channel. Interested readers can refer to references~\cite{naik2018coexistence, burton2009802} for more details on the exponential back-off protocol. 

Values of CSMA/CA or LBE LBT contention parameters vary according to the priority of the packet at the contending device. These parameters include the minimum and maximum values of the Contention Window (CWmin and CWmax, respectively) and the fixed interval for which the channel must be sensed following an ongoing transmission (IFS for Wi-Fi and the defer time (T$_{\textrm{d}}$) for NR-U). The priority of transmissions is determined by the QoS requirements of the application that generates the packet. QoS requirements are specified in terms of the worst-case latency and reliability requirements. A high priority packet has low worst-case latency and high-reliability requirements. Thus, for packets with higher priority, the wait time after the channel becomes idle (i.e., IFS or T$_{\textrm{d}}$) and the CWmin and CWmax values are smaller than those for low priority packets. Wi-Fi and NR-U, both classify packets into four access categories. In Wi-Fi, these categories are referred to as voice (VO), video (VI), best-effort (BE) and background (BK), in decreasing order of priority. 

The contention parameters for the four access categories in Wi-Fi and NR-U are shown in Table~\ref{table:channel-access}. It must be noted from Table~\ref{table:channel-access} that the wait times (i.e., IFS and T$_{\textrm{d}}$) and CW values for NR-U and Wi-Fi packets that belong to a given access category are the same. This ensures that Wi-Fi and NR-U devices are fair to each other in the probability of accessing the channel.

\begin{table*}[htb]
    \centering
    \begin{tabular}{|c|c|c|c|c|c|c|c|c|c|}
    \hline
    \multicolumn{2}{|c|}{\textbf{Access Category}} & \multicolumn{2}{|c|}{\textbf{Wait Time}} & \multicolumn{2}{|c|}{\textbf{CWmin}} & \multicolumn{2}{|c|}{\textbf{CWmax}} & \multicolumn{2}{|c|}{\textbf{TXOP}}\\ \hline
    \textbf{Wi-Fi} & \textbf{NR-U} & \textbf{Wi-Fi (IFS)} & \textbf{NR-U (T$_{\textrm{d}}$)} & \textbf{Wi-Fi} & \textbf{NR-U} & \textbf{Wi-Fi} & \textbf{NR-U} & \textbf{Wi-Fi} & \textbf{NR-U} \\ \hline
    VO & 1 & 25~$\mu$sec & 25~$\mu$sec & 4 & 4 & 8 & 8 & 2.080~msec & 2~msec\\ \hline
    VI & 2 & 25~$\mu$sec & 25~$\mu$sec & 8 & 8 & 16 & 16 & 4.096~msec & 3~msec \\ \hline
    BE & 3 & 43~$\mu$sec & 43~$\mu$sec & 16 & 16 & 1024 & 1024 & 2.528~msec & 8~msec or 10~msec\\ \hline
    BK & 4 & 79~$\mu$sec & 79~$\mu$sec & 16 & 16 & 1024 & 1024 & 2.528~msec & 8~msec or 10~msec\\ \hline
    \end{tabular}
    \caption{Contention Parameters for Wi-Fi and NR-U.}
    \label{table:channel-access}
\end{table*}


\subsubsection{Transmission Duration}
\label{subsubsec:txop}

Contention parameters outlined in the previous subsection control the probability by which contending devices gain access to the channel. Ensuring that this probability is the same for Wi-Fi and NR-U devices, however, does not guarantee fair coexistence between NR-U and Wi-Fi in the shared spectrum. Once Wi-Fi or NR-U devices gain access to the channel, they can transmit uninterrupted (i.e., without contention) for a duration referred to as the transmission opportunity (TXOP). This parameter, again, varies with the access category of packets. Observe from the TXOP values of the different access categories of Wi-Fi and NR-U in Table~\ref{table:channel-access} that NR-U devices can transmit unhindered for a longer duration than Wi-Fi devices. This gives NR-U devices an undue advantage in terms of the average observed throughput. 

It has been previously argued that Wi-Fi and LTE-LAA devices must use the same TXOP values for any given access category in order to ensure air-time fairness. For example, reference~\cite{hp-coex-sc} shows that when both LTE-LAA and Wi-Fi devices coexist, Wi-Fi devices get a smaller share of the channel due to differences in TXOP used by LTE-LAA and Wi-Fi devices. Simulation results in~\cite{ghosh-coex-sc} show that a TXOP duration of 6~msec can achieve fair coexistence between Wi-Fi and NR-U devices. The TXOP values for different access categories to be used in the 6~GHz bands are likely to be standardized by ETSI BRAN. However, it is critical that regardless of the chosen TXOP values, for a given access category, they must be the same for Wi-Fi and NR-U devices to ensure fair coexistence. 

\subsection{Detection of Wi-Fi and NR-U signals}
\label{subsec:coex-mechanism}

\subsubsection{Energy Detection v/s Preamble Detection}
\label{subsubsec:ed-pd}

A key consideration that will govern the efficacy of coexistence between NR-U and Wi-Fi devices in the 6~GHz bands is the choice of the mechanism by which Wi-Fi 6 and NR-U devices will detect each other as well as the corresponding detection threshold. Two possible mechanisms exist for the detection of wireless signals on the air---(i) energy detection, i.e., ED, and (ii) preamble detection (PD). In ED, devices measure the energy locally and determine the channel availability. If the measured energy is greater than a predetermined ED threshold, the channel is declared busy. Otherwise, the channel is considered idle and devices can begin to contend for the channel. PD, on the other hand, requires that devices must transmit and detect a known preamble signal. Upon reception of this preamble, its energy is compared against the PD threshold. The channel is declared busy if the energy is greater than the PD threshold and idle otherwise.

In the 2.4~GHz and 5~GHz bands, Wi-Fi devices use PD to detect other Wi-Fi signals. In order to ensure backward compatibility toward all Wi-Fi generations, IEEE 802.11n/ac/ax devices transmit (and decode) the IEEE 802.11a preamble. On the other hand, ED is used to determine the occupancy of the channel by non-Wi-Fi devices (such as ZigBee in the 2.4~GHz band and LTE-LAA in the 5~GHz bands). Since LTE-LAA was designed at a time when Wi-Fi devices were widely deployed, its designers had two options---(i) adopt the IEEE 802.11a preamble, i.e., transmit this preamble at the beginning of each LTE-LAA frame and decode the IEEE 802.11a preamble to infer channel availability, or (ii) use energy detection to detect both LTE-LAA and non-LTE-LAA signals. Eventually, 3GPP decided to use the latter approach. However, the NR-U work item~\cite{RP-182878} considers extending the detection mechanism beyond ED to include detection of IEEE 802.11a/ax preamble or existing NR signals.

At present, there is no consensus on the choice between ED and PD as the detection mechanism for coexistence between Wi-Fi and NR-U devices in the 6~GHz bands. While some contributions prefer ED~\cite{11-19-1083r1, huawei-coex-sc, 11-19-1088r1, etsi-19-104010r1}, others such as~\cite{cableLabs-coex-sc, quantenna-coex-sc, orange-coex-sc, broadcom-coex-sc} prefer PD. Based on the discussions presented in these contributions, there are several arguments in favor of both mechanisms. 

\begin{table*}[htb]
    \centering
    \begin{tabular}{|p{1.4cm}|p{7.9cm}|p{7.2cm}|}
        \hline
        \textbf{Mechanism} & \textbf{Pros} & \textbf{Cons}  \\ \hline
        Energy & - Simple, low-cost implementation & - No power saving feature \\ 
        Detection & - Technology Neutral & - Not suitable for low detection thresholds \\ 
        & - ED with high threshold can yield improvements through spatial reuse & - ED with high threshold can worsen the hidden node problem \\ \hline
        Preamble & - Reliable at low detection threshold & - More complex implementation than ED \\ 
        Detection & - Power saving through virtual carrier sensing & - Specification of a common preamble will be time-consuming  \\ 
        & - New, efficient common preamble can be defined& - Low threshold in PD exacerbates the exposed node problem \\ \hline
        Hybrid & - Greater flexibility to individual RATs & - Performance in dense scenarios is questionable \\ 
        (ED + PD) & & - Complex implementation \\ \hline
    \end{tabular}
    \caption{A comparison of Energy Detection and Preamble Detection for coexistence between unlicensed RATs.}
\label{tab:ed-pd}
\end{table*}

ED is simple to implement in NR-U and Wi-Fi devices. Additionally, it is argued that ED is technology-neutral in that NR-U (and possible future RATs) need not possess the ability to transmit and decode an arbitrary preamble signal~\cite{etsi-19-104010r1}. PD across different coexisting RATs, on the other hand, necessitates an agreement on which preamble to use. References~\cite{quantenna-coex-sc, orange-coex-sc, broadcom-coex-sc} have argued that the IEEE 802.11a preamble, which is used by all Wi-Fi devices in the 5~GHz bands, is suitable for enabling fair coexistence between NR-U and Wi-Fi devices. This implies that NR-U devices must be equipped with capabilities to transmit and decode 802.11a preambles, while no modifications will be required at 802.11ax devices. It is claimed in~\cite{broadcom-coex-sc} that since typical mobile devices are likely to have access to both NR-U and Wi-Fi modules, 802.11a preambles can be decoded in real-time using the collocated Wi-Fi module. Reference~\cite{att-coex-sc}, on the other hand, argues for a fresh design of a common preamble that can be transmitted and decoded by both NR-U and Wi-Fi devices. The principal merit of this choice is that the 802.11a preamble contains several fields that are unnecessary from the point of view of NR-U and Wi-Fi coexistence. Therefore, a common preamble that contains only the minimum information required to enable coexistence is desirable for maximizing spectral efficiency. 

The advantage of PD lies in the fact that upon decoding the preamble, the duration of the transmission can be inferred. This mechanism, referred to as virtual carrier sensing in Wi-Fi~\cite{burton2009802}, is important in power saving~\cite{broadcom-coex-sc}. Thereafter, the sensing device can enter the sleep mode for the remainder of transmission duration and resume contention for the channel once the channel is known to be idle. Such power saving mechanisms are not possible with ED-based coexistence, where sensing devices must continuously monitor the channel for any activity. In addition to power saving, it is argued in~\cite{broadcom-coex-sc} that for detection thresholds lower than -72~dBm, energy detection is often unreliable, while preamble detection can reliably be used for threshold values up to -82~dBm. 

An alternative to fixed use of ED or PD is to use a hybrid solution, whereby devices use both ED and PD for channel access. This approach is similar to how Wi-Fi currently operates in the 5~GHz bands---detect Wi-Fi signals at -82~dBm and non-Wi-Fi signals at -62~dBm. Such schemes have also been proposed for LTE-LAA devices operating in the 5~GHz bands~\cite{hong2019lightweight}. In the 6~GHz bands, this would extend as follows: Wi-Fi and NR-U devices sense inter-RAT transmissions using ED, which is technology-neutral. However, intra-RAT transmissions can be detected using technology-specific preambles, i.e., PD. The benefit of this approach is that individual RATs can optimize their performance by adapting spatial reuse through dynamic adaptation of the PD threshold. This spatial reuse technique is widely considered in IEEE 802.11ax~\cite{11axDraft}. However, contributions against this proposal, such as~\cite{etsi-19-104010r1}, have argued that the sole benefit of using this hybrid detection approach, i.e., flexibility to individual RATs and resulting performance gains, is questionable in dense deployment scenarios. 

The merits and demerits of ED, PD and a hybrid approach are outlined in Table~\ref{tab:ed-pd}.
\subsubsection{Choice of the Detection Threshold}
\label{subsubsec:threshold}

The detection threshold used by Wi-Fi and NR-U devices to detect and defer to each other is one of the most contentious topics being debated at the IEEE Coexistence SC and ETSI BRAN meetings~\cite{11-19-2150r6}. Wi-Fi devices in the 2.4~GHz and 5~GHz bands use different detection thresholds to detect other Wi-Fi and non-Wi-Fi signals. The former threshold is set to -82~dBm, while the latter is -62~dBm. During LTE-LAA design, there was considerable debate over the choice of detection threshold used by LTE-LAA devices to detect Wi-Fi signals. This was a contentious issue because of two reasons, (i) at the time of LTE-LAA design Wi-Fi devices were already ubiquitous in home and enterprise wireless networks with millions of users worldwide; on the other hand, LTE-LAA was a new technology that could potentially negatively affect the performance of existing Wi-Fi users, and (ii) it was not possible to change the detection thresholds across the millions of already deployed Wi-Fi APs and STAs so that LTE-LAA and Wi-Fi could detect each other using the same threshold. Eventually, it was decided that LTE-LAA devices would use a fixed energy detection threshold (-72~dBm) to detect both LTE-LAA and non-LAA signals~\cite{11-19-1083r1}. 

The argument used while adopting the -72~dBm threshold for LTE-LAA devices in the 5~GHz bands is no longer valid in the greenfield 6~GHz bands. Furthermore, the first two technologies to operate in these bands on an unlicensed basis, i.e., IEEE 802.11ax and NR-U, are still undergoing (albeit final stages of) design. Thus, the detection threshold for the 6~GHz bands can be decided after thoroughly studying its impact on the performance of Wi-Fi and NR-U networks, especially taking into consideration the nature of applications to be supported by these RATs.

Detection mechanisms and thresholds used in the 5~GHz bands are known to be unfair toward one technology or the other (based on the network topology). However, some contributions argue Wi-Fi and NR-U operations in the 6~GHz bands must adopt the \textit{status quo} in terms of these mechanisms/thresholds from the 5~GHz sharing rules~\cite{etsi-19-103023}. Nevertheless, a vast majority of contributions to the IEEE Coexistence SC and ETSI BRAN have argued for a technology-neutral and common threshold to be used in the 6~GHz bands~\cite{11-19-1083r1, ghosh-coex-sc, 11-19-1088r1, etsi-19-103016}. Despite this, however, there is no consensus on the choice of an appropriate detection threshold. Reference~\cite{ghosh-coex-sc}, for example, argues that a common ED threshold of -82~dBm improves the performance of both Wi-Fi and 3GPP-based networks. However, a different common threshold (-62~dBm) is shown to be more effective for both NR-U and Wi-Fi performance in~\cite{huawei-coex-sc}, while references~\cite{broadcom-coex-sc, etsi-19-103016, etsi-19-104010r1} argue for the use of -72~dBm. 

In addition to the aforementioned references that advocate fixed thresholds, some contributions have argued for adjustable threshold values so that interference can be mitigated in real-world deployment scenarios~\cite{orange-coex-sc}. Reference~\cite{broadcom-coex-sc} proposes a hybrid solution where a common detection threshold is used across different RATs (i.e., for NR-U's detection of Wi-Fi signals and vice-versa). Each RAT can then use adaptive thresholds for intra-RAT detection (e.g., detection of Wi-Fi signals at a Wi-Fi transmitter) to maximize their respective performance using spatial reuse. In this approach, a lower threshold (e.g., -82~dBm) could be used to detect intra-RAT transmissions, while a higher threshold (e.g., -62/-72~dBm) can be used for detecting inter-RAT (i.e., Wi-Fi detection at NR-U transmitters and vice-versa) transmissions.
 
The references mentioned above present their arguments with an underlying assumption that NR-U and Wi-Fi devices both use LBT-based channel access protocols. However, Wi-Fi 6 devices can also use MU-OFDMA for scheduled uplink and downlink transmissions. In this case, although the AP must first sense and contend for the channel for transmitting the scheduling Trigger Frame, uplink transmitter STAs do not need to contend for the channel. Furthermore, the presence of more than one active transmitter at a time (as is the case in uplink MU-OFDMA transmissions) reduces the probability of transmitting devices being hidden to sensing devices. Thus, the implications of a particular detection threshold on the performance of Wi-Fi 6 and NR-U in such coexistence scenarios could be different in comparison to single user Wi-Fi transmissions as noted in~\cite{11-19-1111r0}.

%% file: adj-channel.tex
\section{Adjacent Channel Interference Issues}
\label{sec:adjacent}

\begin{figure}[htb]
 \centering
 \includegraphics[scale=0.4]{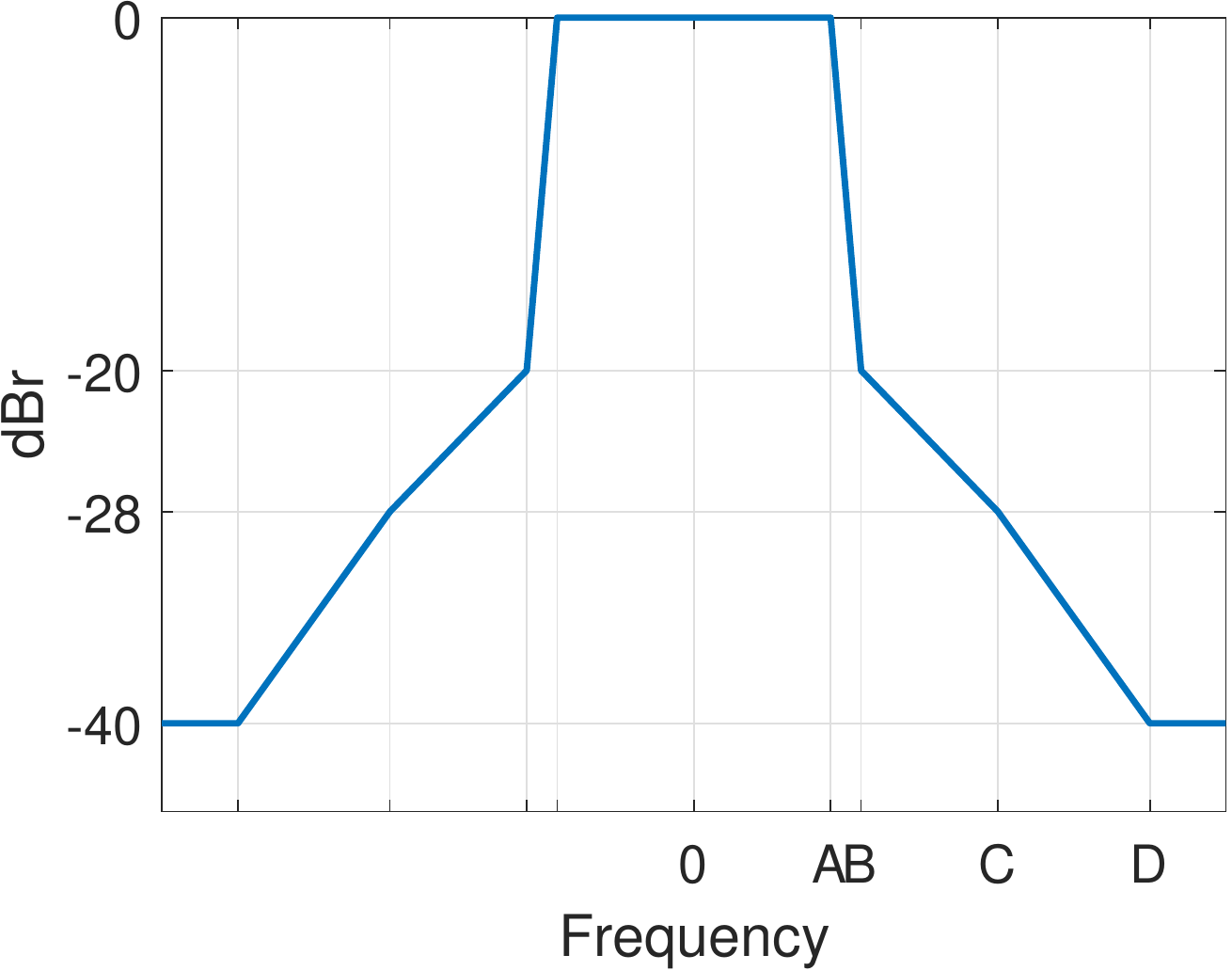}
 \caption{The default (class A) spectral mask for Wi-Fi.}
 \label{fig:spectral-mask}
\end{figure}

When a wireless device radiates, a small fraction of its radiated power \textit{leaks} on to adjacent channels. This is true for any wireless device, although the extent to which a device leaks this power on to adjacent channels depends on the device type, device cost, the spectrum band of operation, and the regional regulations. For every geographical region, the regulatory agencies require conformance of out-of-band emission characteristics to specific standards. For example, in the US, the FCC requires that Wi-Fi devices operating in the 5~GHz bands adhere to out-of-band emissions shown in Fig.~\ref{fig:spectral-mask}. The emission characteristics shown in Fig.~\ref{fig:spectral-mask} are referred to as the spectral mask. Specifically, Table~\ref{table:abcd} refers to the \textit{class A} spectral mask for Wi-Fi devices operating in unlicensed bands~\cite{classD}. The frequency separation between the center frequency and points A, B, C, and D shown in Fig.~\ref{fig:spectral-mask} for the class A mask is given by Table~\ref{table:abcd}. 

\begin{table}[htb]
 \centering
 \caption{Frequency offsets A, B, C and D for the class A mask.}
 \label{table:abcd}
 \begin{tabular}{|c|c|c|c|c|}
  \hline
  \scriptsize{\textbf{Bandwidth}} & \scriptsize{\textbf{A (0 dBr)}} & \scriptsize{\textbf{B (-20 dBr)}} & \scriptsize{\textbf{C (-28 dBr)}} & \scriptsize{\textbf{D (-40 dBr)}} \\ \hline
  $20$~MHz & $9$~MHz & $11$~MHz & $20$~MHz & $30$~MHz \\ \hline
  $40$~MHz & $19$~MHz & $21$~MHz & $40$~MHz & $60$~MHz \\ \hline
  $80$~MHz & $39$~MHz & $41$~MHz & $80$~MHz & $120$~MHz \\ \hline
  $160$~MHz & $79$~MHz & $81$~MHz & $160$~MHz & $240$~MHz \\ \hline
 \end{tabular}
\end{table}

Devices conforming to regional regulations must radiate all their power within the required spectral mask (such as the one shown in Fig.~\ref{fig:spectral-mask} and specified by Table~\ref{table:abcd}). However, different commercial devices have different spectral mask characteristics. In general, lower the cost of the device more is the radiated power outside the desired channel (i.e., outside point A). Higher-end devices, on the other hand, possess superior radio frequency filters, thereby radiating most of their power within the desired channel. 

As noted in Sec.~\ref{subsec:adjacent}, at the lower edge of the U-NII-5 band lies the 5.9~GHz band, which is reserved in the US and Europe for ITS applications. Since the ITS band enables many vehicular safety applications, it is necessary to ensure that interference generated from U-NII-5 band unlicensed devices remains minimal at ITS band receivers. The authors in~\cite{pang2018sophisticated} report the performance of a DSRC-based V2X communication system in the presence of Wi-Fi interferers when DSRC and Wi-Fi devices operate in channels 180 and 177\footnote{Channel 177 is a 20~MHz Wi-Fi channel in the U-NII-4 band. See~\cite{naik2019impact}.}, respectively. The authors use a metric referred to as the safety alert failure rate, which signifies the delivery rate of \textit{critical} basic safety messages. A basic safety message is referred to as critical at a given receiver if the sender of the message and the given receiver are at risk of a collision. The authors report that the performance of such critical messages is practically unaffected by the presence of Wi-Fi devices in adjacent channels. 

On the other hand, it is reported in  reference~\cite{naik2019impact} that if Wi-Fi devices in the U-NII-5 band are deployed in outdoor environments, where the Wi-Fi and C-V2X devices can be placed within a small distance from each other, the performance of C-V2X devices is significantly affected. This is especially true if the distance between the transmitting C-V2X device and its receiver is large (i.e., when the SINR is small). This performance loss is captured in terms of the decline in Packet Delivery Ratio of C-V2X basic safety messages when Wi-Fi devices are nearly 10~m away from C-V2X receivers in an urban setting~\cite{ltebasedv2x}. Experiments performed by the 5G Automotive Association in real-world settings~\cite{ex-parte-5gaa} further show that unlicensed devices operating in the lowermost channels of the U-NII-5 band can result in interference levels of -27~dBm/MHz, which is unacceptable in terms of performance degradation at C-V2X receivers.

%% file: challenges.tex
\section{Research Challenges}
\label{sec:challenges}

\subsection{Coexistence between incumbents and unlicensed devices}
The introduction of new unlicensed RATs in a band introduces several challenges in achieving coexistence between the band's incumbent users and unlicensed devices. The 6~GHz bands are no exception to this, as highlighted in Sec.~\ref{sec:incumbent}. Several incumbent users have highlighted the need for Wi-Fi and NR-U devices to exercise more caution during their operation in these bands. This is critical because once the proliferation of unlicensed devices takes place in the 6~GHz bands, much like in the 5~GHz bands, it becomes extremely difficult to identify and enforce restrictions on potentially rogue unlicensed devices that are already deployed. The following aspects of coexistence between incumbent and unlicensed devices must be investigated.

\subsubsection{Efficiency of the AFC System} 
In the US, incumbent users in the U-NII-5 and U-NII-7 bands will be protected by using the concept of exclusion zones. Unlicensed devices operating in these bands must connect to the AFC system database to determine if their location is within the exclusion zone of any incumbent receiver. If an unlicensed device is found to be within the exclusion zone of an incumbent receiver, it can operate on all frequencies other than the operating frequency of the incumbent receiver. Such database-driven coexistence mechanisms are shown to be effective in other frequency bands~\cite{bhattarai2016overview}. Furthermore, proponents of unlicensed services have claimed that unlicensed devices operating in the U-NII-5/7 bands are highly unlikely to interfere with incumbents operating in these bands. Even though these claims are backed by preliminary simulation studies and experimental results, considering the stringent outage constraints placed on incumbent services, these coexistence scenarios must be investigated with rigorous analytical models and experimental studies. 

At the same time, the AFC system must not be too conservative in allowing unlicensed access in the U-NII-5 and U-NII-7 bands. The boundary of a conventional exclusion zone is determined based on an estimate of the union of all likely interference scenarios, and this results in an overly conservative boundary that often leads to low spectrum utilization efficiency.  To address this shortcoming, authors in~\cite{bhattarai2015defining} have proposed the concept of Dynamic Incumbent Protection Zones (DIPZ). Unlike conventional exclusion zones, a DIPZ is a dynamic spatial separation region surrounding an incumbent receiver whose boundary is dynamic in terms of geolocation, time, and frequency. Such schemes have the potential to increase spectrum utilization efficiency while providing protection to the incumbent users. More research is needed to determine whether such schemes can be adopted in the 6 GHz bands to improve spectrum utilization efficiency.

\subsubsection{Interference characterization of VLP and LPI devices} In comparison to incumbent receivers operating in the U-NII-5/7 bands, the nomadic incumbent users in the U-NII-6/8 bands can be more vulnerable to interference from unlicensed devices. This is because receivers of such nomadic incumbent systems are more likely to come in close proximity of unlicensed devices. To tackle this issue, the FCC R\&O restricts unlicensed operations in the U-NII-6/8 bands to indoor-only LPI devices, thereby relying on lower transmit powers and building entry losses to mitigate interference at incumbent receivers. Although the FCC believes such measures are sufficient to protect the incumbent users in these bands, several incumbent users have raised concerns regarding interference from LPI devices (see Sec.~\ref{subsec:6-8}). Similar issues can be expected if VLP devices are permitted to operate in the U-NII-6/8 bands, as indicated in the FCC FNPRM. To ensure that the performance of incumbent systems remains uncompromised, it is necessary to characterize the interference resulting from VLP and LPI devices at incumbent receivers, especially before unlicensed LPI and VLP devices proliferate in these bands.

Interference characterization of LPI and VLP devices can be performed by considering the typical operating characteristics of unlicensed and incumbent systems. This includes factors such as the link budget of incumbent systems, operating environments (such as indoor/outdoor stadiums, public parks, concert halls, etc.), locations of incumbent receivers and unlicensed transmitters, transmit power of unlicensed devices, appropriate propagation models, etc. For example, given that a special event occurs inside an indoor stadium, the locations of incumbent and unlicensed devices can be modeled as two independent 2-D point processes~\cite{baccelli2010stochastic}. Using an appropriate propagation model, such as the WINNER model~\cite{meinila2010d5}, the impact of VLP and LPI devices can be analyzed as the fraction of time during which the unlicensed interference power at an arbitrary incumbent receiver exceeds the link budget of the system.

\subsubsection{Mechanisms for detection of incumbent users in the U-NII-6 and U-NII-8 bands} 

Without a method for detecting incumbents and/or coordinating channel access, access by unlicensed devices needs to be strictly limited.  For instance, the FCC FNPRM proposes to limit the transmission power of unlicensed devices operating outdoors (such as the case with VLP devices) while the FCC R\&O only allows indoor operations (such as the case with LPI devices). However, such restrictions may significantly limit the utility of the U-NII-6/8 bands to unlicensed users.

An efficient way to ensure that incumbent users remain interference-free while unlicensed devices have the maximum operational flexibility is to define incumbent user detection mechanisms for unlicensed devices. If any incumbent user activity is detected in the vicinity of the monitoring unlicensed device, it must cease its operations on that channel and move to other frequencies. This is similar in spirit to the Dynamic Frequency Selection (DFS) mechanism deployed at Wi-Fi devices operating in the U-NII-2C band~\cite{naik2018coexistence}. However, the DFS mechanism is known to be conservative beyond necessity and makes Wi-Fi operations infeasible in channels where radar systems operate. Thus, alternate mechanisms that are more efficient at utilizing the spectrum must be defined in the 6~GHz bands. Such a dynamic frequency coordination mechanism can leverage the signal characteristics of incumbent users and detect their on-air presence. 

To define such a dynamic frequency coordination mechanism, the following specific aspects must be studied --- (i) what characteristics of the incumbent signal can be used for reliable detection? (ii) in order to alleviate the hidden node problem, what threshold energy level must the unlicensed devices use for detecting the presence of incumbent users? (iii) what should the initial sensing time be for which unlicensed devices \textit{listen} without transmitting on the channel?, (iv) if incumbent activity is not detected, how often must unlicensed devices perform the detection routine? (v) once incumbent user activity is detected, what is the maximum time for which unlicensed devices can operate on the channel without degrading the incumbent system performance? and (vi) how soon can unlicensed devices resume operating on the channel if incumbent activity is detected? 

\subsubsection{Enforcement of rogue unlicensed devices} Despite the use of interference mitigation schemes like DFS, weather radar operators in the 5~GHz bands have reported harmful interference from Wi-Fi operations~\cite{radarinterference1, radarinterference4, radarinterference5}. Once deployed, the enforcement of such rogue Wi-Fi transmitters is known to be an extremely difficult problem~\cite{naik2018coexistence}. Similar problems can occur in the 6~GHz bands. Standard power APs operating in the U-NII-5 and U-NII-7 bands are required to provide their FCC identifier during their registration with the AFC system database~\cite{fcc-ro}. The enforcement of such standard power APs, therefore, is likely to be straightforward. However, recall that LPI unlicensed devices are permitted to operate inside the exclusion zones of incumbent receivers in the U-NII-5 and U-NII-7 bands as long as they operate indoors. If building entry losses are small, such devices can interfere with incumbent users.
Further, as proposed in the FCC FNPRM~\cite{fcc-ro}, if VLP devices are permitted to operate in the 6~GHz bands, incumbent receivers (especially those in the U-NII-6 and U-NII-8 bands) can be susceptible to interference from nearby VLP devices. 

In order to ensure that the performance of incumbent users is not compromised, enforcement of sharing rules at potentially rogue unlicensed devices must be proactively addressed in these bands. Spectrum sharing enforcement involves the following three stages, (i) identification of interfering device, (ii) localization of the interfering device, and (iii) imposing punitive actions. The state-of-the-art literature on these stages of enforcement (such as~\cite{park2014security, kumar2014blind, kumar2016phy, tan2011cryptographic, yang2012enforcing, kumar2018enforcement, kumar2019crowd, awad2017rogue, le2012rogue, zheng2014accurate, khaledi2017simultaneous, zhou2017detecting, wang2016locating}) must be investigated for their suitability for the 6~GHz bands. If these mechanisms are deemed inappropriate, novel mechanisms that consider the specific characteristics of incumbent and unlicensed users of the 6~GHz bands must be developed.

\subsection{Unlicensed RAT Optimization Issues}
\subsubsection{Fairness of IEEE 802.11be with legacy IEEE 802.11 devices}
As noted in Sec.~\ref{subsec:throughput}, synchronous multi-primary MLA devices can often gain a \textit{free-ride} on the secondary link. In a two-link scenario, recall that such MLA-capable devices contend for access on both links. If the MLA-capable device gains access to one channel (say channel 1), it performs an ED check on the other channel (say channel 2). Channels 1 and 2 are aggregated if channel 2 is idle, i.e., if the ED check passes. Note, however, that the MLA-capable device gained access to channel 2 without having to undergo the entire contention mechanism (i.e., IFS + back-off). As noted in Sec.~\ref{subsec:throughput}, this free-riding is beneficial to the MLA-capable device in terms of increasing its observed throughput. However, all single link devices (including legacy Wi-Fi devices) on channel 2 must wait until their respective back-off counters expire. Thus, this synchronous multi-primary MLA characteristic is unfair toward the single link Wi-Fi devices operating on both channels. 

What makes the unfairness in the above example worse is that the default contention engine in Wi-Fi resets the CW (to CWmin) on channel 2 if the packet transmitted on channel 2 is successfully delivered~\cite{11-19-1505r2}. However, transmission on channel 2 was initiated due to MLA-enabled aggregation. It must be noted that this behavior can have a negative impact not only on single Wi-Fi (legacy and 802.11be) devices but also on NR-U devices contending on either channels. 

Some preliminary approaches to mitigate this unfairness include: (i) resetting the CW only on that channel which won access to the channel through fair contention, i.e., the channel on which the MLA-capable device counted its back-off counter down to zero (channel 1 in the above example) and not on other channel(s) even if the transmission is successful (channel 2 in the above example), (ii) compensate for the free-riding on one (or more) channels by picking a new back-off counter for that channel (channel 2 in the above example)~\cite{11-19-1633r1}, (iii) if the selection of a new back-off counter with the previous CW value does not alleviate the unfairness, then the CW can be doubled and a new back-off counter can be chosen~\cite{11-19-1633r1}, and (iv) aggregate channels only if the transmission duration on the channel(s) is smaller than a certain threshold time. This approach can ensure that MLA-capable devices use synchronous multi-primary MLA only for real-time applications---which are characterized by shorter packet sizes than other traffic categories~\cite{11-19-0065r6}---but not for packets belonging to other traffic categories.

\subsubsection{Simultaneous Transmit \& Receive Constraints in IEEE 802.11be}
\begin{figure}[htb]
    \centering
    \includegraphics[trim={7.2cm 6.5cm 10.2cm 15cm},clip,scale=0.8,angle=-90]{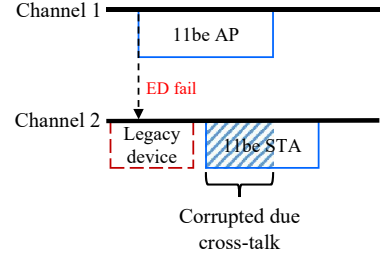}
    \caption{The simultaneous transmit-receive (STR) constraint during MLA in IEEE 802.11be.}
    \label{fig:str}
\end{figure}

In both synchronous MLA schemes, i.e., synchronous single primary and multi-primary MLA, there can arise a scenario where the MLA-capable AP is transmitting on one channel (channel 1 in Fig.~\ref{fig:str}) while an STA starts transmitting a packet to the AP on the other channel (channel 2 in Fig.~\ref{fig:str}). However, if the two channels are such that there is an inter-link cross-talk at the AP, the AP may be unable to decode the packet received from the STA on channel 2. This is referred to as the simultaneous transmit-receive (STR) constraint~\cite{11-19-1405r7, 11-19-1547r3}. The STR constraint presents a significant challenge in the operation of synchronous MLA schemes. Since synchronous MLA is likely to be the only MLA scheme suitable for aggregating channels in the 5~GHz and 6~GHz bands, effective solutions to avoid the STR constraint are required. At the time of writing this paper, the IEEE 802.11be Task Group is actively working toward solutions that prevent the occurrence of the STR problem in synchronous MLA schemes.

\subsubsection{FBE Mechanisms for NR-U}
\label{subsubsec:fbe-nru}
Unlike IEEE 802.11be, specification work on NR-U has emphasized the design of channel access protocols for FBE LBT. However, FBE LBT in NR-U is targeted for scenarios where the absence of Wi-Fi networks can be guaranteed. One such setting is a factory environment where a single operator manages the deployment of all unlicensed RATs and hence, can guarantee an environment ``clean'' of Wi-Fi~\cite{R1-1912938}. This is certainly possible in the greenfield 6~GHz bands. However, certain challenges must be addressed to ensure smooth FBE LBT operations. For example, current ETSI regulations recommend one-shot LBT for FBE~\cite{etsi-5GHz}. This implies that an FBE device senses the channel without random back-off (i.e., category 1/2 LBT), and if the channel is detected as idle, the device transmits. This can lead to persistent collisions as all devices that have packets to transmit sense and start transmitting at the same time~\cite{R1-1911822}. To mitigate this issue, several solutions have been proposed by company contributions. References~\cite{R1-1911822, R1-1912449} propose to introduce randomness in the sensing mechanism, essentially leading to FBE operations shown in Fig.~\ref{fig:fbe}, i.e., Category 3/4 LBT at the beginning of each frame interval. On the other hand, references~\cite{R1-1912012, R1-1912449} propose to use various other schemes such as coordinated transmissions, directional sensing and transmissions to mitigate interference during simultaneous transmissions. 

At present, the above proposals are only under consideration and it is likely that the 3GPP will adopt the status-quo from the 5~GHz bands moving into the 6~GHz bands. However, before unlicensed devices proliferate in the 6~GHz bands, there is an opportunity to optimize the NR-U-only (i.e., regions where the absence of Wi-Fi operations can be guaranteed) FBE LBT operations by considering the above mechanisms as the starting point. The prominent research directions include investigations on: (i) contention time, i.e., should the devices contend for the frame interval at the beginning of each frame or at the end? (ii) contention type, i.e., what LBT type should be used during the contention interval (Category 2/3/4)? (iii) interference coordination mechanisms, i.e., should category 2 LBT be adopted during the contention interval, are directional sensing and transmissions effective?

\subsection{Coexistence between Unlicensed Technologies}

\subsubsection{Performance Analysis of FBE LBT in the 6~GHz Bands}
If FBE LBT is to be used across different unlicensed RATs operating in the 6~GHz bands, a global clock is required. This is likely to be, in general, more readily achievable for NR-U networks since the NR-U gNB is likely to be integrated with the cellular network. However, to justify this added requirement in Wi-Fi networks, a through investigation is required on the performance benefits, if any, of using FBE LBT across all unlicensed RATs in the 6~GHz bands. 

Assuming that the NR-U FBE LBT protocol is based on category 3/4 LBT (as discussed above in Sec.~\ref{subsubsec:fbe-nru}) and Wi-Fi systems adopt a similar FBE LBT protocol, gains resulting from a decrease in the number of hidden nodes (as claimed in reference~\cite{ex-parte-qualcomm} and discussed in Sec.~\ref{subsubsec:lbe-fbe}), and its impact on the system performance in dense settings must be evaluated. Furthermore, since the draft standard of IEEE 802.11ax is in its final stages of completion, Wi-Fi 6 devices are likely to use LBE LBT in the 6~GHz bands at least during the initial stages of deployment. This issue is acknowledged in~\cite{qualcomm-sync}, where it is argued that although not all unlicensed devices can be required to conform to the synchronized FBE access procedure, the larger the number of devices that conform to FBE LBT, the better it is for overall system performance. 

It must be noted that the adoption of FBE LBT in the 6~GHz bands is only in the proposal stage. As discussed in Sec.~\ref{subsubsec:lbe-fbe}, FBE LBT was discussed as a candidate channel access mechanism during the design of LTE-LAA. The primary reason for the adoption of LBE LBT in LTE-LAA fairness is---LBE LBT devices gain a higher share of the channel when coexisting with FBE LBT devices. Thus, notwithstanding the performance gains of FBE LBT, the challenges it introduces in terms of coexistence with LBE LBT devices remains an open problem. Some preliminary coexistence mechanisms between FBE and LBE LBT have been proposed in~\cite{qualcomm-sync, R1-1912088}. Before FBE LBT-based access to 6~GHz bands is considered, the efficacy of existing coexistence mechanisms (such as those in references~\cite{qualcomm-sync, R1-1912088}) must be investigated or novel coexistence mechanisms must proposed and evaluated.

\subsubsection{Analysis and Optimization of Channel Access Parameters}
Traditionally, Wi-Fi networks have been unable to satisfy the QoS requirements of high reliability and latency-sensitive applications. This inability stems in part from the unlicensed (and hence, non-guaranteed) nature of the wireless spectrum on which Wi-Fi networks operate. However, the fundamental cause of this problem is the tail (i.e., worst-case) access latency encountered due to Wi-Fi's CSMA/CA-based channel access procedure~\cite{chatzimisios2002throughput, sakurai2007mac}. With the likely introduction of MLA in IEEE 802.11be and the availability of a large number of \textit{links} in the 6~GHz bands to contend on, the tail latency in IEEE 802.11be is likely to be substantially lowered (as discussed in Sec.~\ref{subsec:latency}). A significant research problem, therefore, is to quantitatively assess the latency gains of MLA in IEEE 802.11be over traditional single-link Wi-Fi networks. There is rich literature on the latency and throughput analysis of IEEE 802.11 networks (e.g.,~\cite{chatzimisios2002throughput, sakurai2007mac, bianchi2000performance, dai2012unified, huang2007throughput, xu2008access}, and the references therein). A vast majority of these works rely on one-dimensional or multi-dimensional Markov Chain models of the Wi-Fi MAC protocol. Naturally, such models can be extended to study the behavior of MLA in different traffic conditions. Such studies can provide an accurate assessment on whether future Wi-Fi devices (with features like MLA) can reliably enable the emerging QoS-sensitive applications in the 6~GHz bands.

An additional research problem that will likely gain traction in the upcoming years is the optimization of the channel access parameters used by the LBT-based MAC protocols in NR-U and Wi-Fi. At the time of writing this paper, the contention parameters outlined in Table~\ref{table:channel-access} are likely to be reused in the 6~GHz bands. Here, the problem arises in the coexistence of the emerging QoS-sensitive applications (such as AR/VR) with the existing ones (such as voice and video traffic). Whether the emerging class of wireless applications can be supported reliably using traffic classes that were originally designed for voice and video traffic remains to be seen. If the channel access parameters for these existing traffic classes are deemed insufficient in supporting such applications, like several other aspects in the 6~GHz bands, channel access parameters can be optimized without encountering backward compatibility hurdles. If this were to be true, the Markov Chain-based analytical models (as noted above~\cite{chatzimisios2002throughput, sakurai2007mac, bianchi2000performance, dai2012unified, huang2007throughput, xu2008access}) can be used in the optimization of 6~GHz bands-specific channel access parameters.

\subsubsection{Detection Mechanism and Threshold for NR-U and Wi-Fi Coexistence} 
Perhaps the most significant parameter influencing the coexistence performance of NR-U and Wi-Fi is the detection threshold used by Wi-Fi and NR-U devices to detect and back-off to each other. While this subject has been widely discussed (as discussed in Sec.~\ref{subsubsec:threshold}), there is no consensus on the choice of an appropriate value for the detection threshold. The chosen detection threshold not only has a direct impact on the performance of the two RATs when they coexist but also has implications on which mechanism (i.e., energy detection or preamble detection) is eventually chosen by NR-U and Wi-Fi devices to detect each other. For example, if the detection threshold selected is too low (such as -82~dBm), the only suitable mechanism by which Wi-Fi and NR-U devices can detect each other is preamble detection. On the other hand, if the chosen detection threshold is higher (such as -72~dBm or -62~dBm), energy detection can be reliably used for NR-U and Wi-Fi signal detection.

If preamble detection is used as the detection mechanism in the 6~GHz bands, an additional problem emerges---a suitable common preamble must be chosen for the 6~GHz bands. The criteria for the design of such a preamble are: (i) The preamble must (ideally) convey information on the duration of the following packet. This, as discussed in Sec.~\ref{subsec:coex-mechanism}, can increase the energy efficiency of devices by enabling devices to \textit{sleep} for the remainder of the packet duration once the duration is decoded; (ii) The accuracy of detection of the preamble must be high across both technologies, i.e., NR-U and Wi-Fi; (iii) The complexity of detection of the preamble at Wi-Fi and NR-U devices, given their dissimilar and non-compatible PHY layers, must be low; (iv) The preamble must be spectrally efficient, i.e., it must consume the minimum possible amount of resources while carrying the minimum necessary information for enabling efficient coexistence. This implies that technology specific information, such as those present in the IEEE 802.11 preamble, must be kept at a minimum while designing a new common preamble. (v) In an ideal scenario, the preamble must be forward looking, i.e., the aforementioned factors must also hold for a potentially new RAT that may operate in the 6~GHz bands in the future.

In order to determine a suitable detection threshold that enables fair and efficient coexistence between NR-U and Wi-Fi, the impact of the chosen threshold on the system-wide performance of the two RATs must be studied. In doing so, it must be noted that small-scale experimental studies can often lead to misleading conclusions~\cite{mehrnoush2018analytical}. Thus, the coexistence of NR-U and Wi-Fi must be studied using rigorous and extensive analytical models, simulation platforms, and large-scale experiments. The foundation for such studies has already been laid out during the study of LAA--Wi-Fi coexistence in the 5~GHz bands~\cite{mehrnoush2018analytical, li2016modeling, ajami2017modeling}. Such analytical models, supplemented by experimental and simulation results, must be extended to study the 6~GHz coexistence problem.

\subsubsection{Impact of MU-OFDMA on NR-U and Wi-Fi Coexistence}
A key distinguishing factor between the LAA--Wi-Fi and the NR-U--Wi-Fi coexistence scenarios is that during the former study, 802.11ac was the default Wi-Fi standard under consideration. Today, on the other hand, Wi-Fi 6 is the new and upcoming Wi-Fi standard, which introduces MU-OFDMA for increased MAC layer efficiency. The use of MU-OFDMA in Wi-Fi 6, especially in the uplink, introduces several similarities in the operations of Wi-Fi and NR-U. For example, for downlink transmissions, the NR-U gNB and the Wi-Fi 6 AP contend for channel access and transmit packets to the designated UEs/STAs on specific RBs/RUs. On the other hand, for uplink transmissions, the NR-U gNB/Wi-Fi 6 AP will first contend for the channel and schedule a certain number of RBs/RUs to specific UEs/STAs. OFDMA-based downlink transmissions in Wi-Fi are similar to CSMA/CA-based single user transmissions in that the Wi-Fi AP contends for the channel and occupies the entire channel bandwidth. However, in the case of uplink OFDMA transmissions, there are likely to be more than one active transmitters transmitting simultaneously on orthogonal frequency resources. As a result, for a given detection threshold, the probability that at least one of the uplink Wi-Fi transmitters is within the sensing range of NR-U devices increases, and the probability of Wi-Fi devices being hidden to NR-U devices diminishes. Furthermore, transmit power control, which was optionally used in legacy Wi-Fi systems is introduced as a mandatory feature for uplink MU-OFDMA transmissions in Wi-Fi 6. As a result, the use of MU-OFDMA in Wi-Fi, especially for uplink transmissions, is likely to have different implications on the choice of the optimal detection threshold. 

From the above discussions, it is clear that the impact of MU-OFDMA on NR-U--Wi-Fi coexistence is another critical subject worthy of investigation. Intuitively, it is expected that the reduced hidden node probability due to a greater number of simultaneously active transmitters will improve the system performance of both Wi-Fi and NR-U. This intuition must be verified using rigorous models, and simulation and experimental studies. Again, the analytical models developed for LAA--Wi-Fi coexistence can serve as the starting point for NR-U--Wi-Fi coexistence. For example, reference~\cite{ajami2017modeling} studies the coexistence performance of LAA and 802.11ax for both single user and MU-OFDMA transmissions. Such models can be extended to study the 6~GHz coexistence problem. Furthermore, if MU-OFDMA is found to be more suitable in promoting fair and efficient coexistence in the 6~GHz bands, the 6~GHz bands provide a unique opportunity where single user transmissions can indeed be disabled (since there are no legacy Wi-Fi operations in these bands).


\subsection{Adjacent Channel Interference}
Unlicensed devices operating in the lowermost channels of the U-NII-5 band can potentially interfere with and degrade the performance of ITS band technologies. In the US, the FCC has proposed to allow exclusive C-V2X operations in channel 183---the uppermost 20~MHz channel in the ITS band~\cite{59ghz-nprm}. The FCC R\&O does not permit unlicensed operations in moving vehicles and trains in any of the four 6~GHz bands. This eliminates in-vehicle coexistence scenarios, where unlicensed devices and C-V2X radios operate in close proximity~\cite{dotTestPlan}. However, if the AFC system determines that a given location does not lie within exclusion zones of any U-NII-5 incumbent receivers, unlicensed deployments can be permitted in outdoor environments. Such use-cases include Wi-Fi deployments at cafes and restaurants in urban areas~\cite{dotTestPlan}. C-V2X and future NR V2X devices remain susceptible to interference from such roadside unlicensed deployments. 

As discussed in Sec.~\ref{sec:adjacent}, the subject of adjacent channel interference has been investigated in the literature. However, there is no consensus on whether such adjacent channel interference can significantly impact the performance of applications enabled by V2X communications. Consequently, there is an urgent need to demonstrate the ability (or lack thereof) of unlicensed devices to operate in the lower channels of the U-NII-5 band without affecting the performance of V2X RATs. Such a study would require (i) the characterization of the out-of-band emissions of unlicensed devices (in comparison to the characteristics mandated by regional regulators, such as~\cite{classD}), (ii) realization of accurate deployment characteristics in simulation environments and test-bed experiments, including the density and locations of (V2X and unlicensed) devices, typical transmit powers of (V2X and unlicensed) transmitters, propagation models that reflect real-world scenarios, practical message sizes for V2X devices, and realistic duty cycles for unlicensed transmitters, (iii) analysis of the impact of interference on link-level metrics (i.e., SINR to block error rate curves), system-level metrics (such as PDR), and application-level metrics. Further, in addition to analyzing the impact of adjacent channel interference on day-1 safety applications, its impact on advanced vehicular safety applications---such as those supported by NR V2X---must also be investigated.

%% file: conclusions.tex
\section{Conclusions}
\label{sec:conclusions}

In this paper, we summarize key points from our comprehensive survey on issues surrounding the opening up of 6~GHz bands for unlicensed access in the US and Europe. We describe the main features and mechanisms that leverage the abundance of spectrum available in these bands in supporting emerging applications such as wireless AR/VR and mobile gaming. Furthermore, we elaborate on the challenges encountered in the coexistence between unlicensed RATs and incumbent users of the band. We describe several concerns raised by the incumbent users as well as defenses provided by the proponents of unlicensed services. We then discuss the challenges that are likely to be encountered when NR-U and Wi-Fi devices operate and coexist in the 6~GHz bands. In doing so, we highlight the key differentiating factor of the 6~GHz bands, where all unlicensed RATs will be allowed to operate for the first time and where coexistence mechanisms can be re-designed with careful consideration. We summarize the key research challenges that need to be addressed in order to enable efficient, fair and harmonious coexistence among unlicensed users as well as between unlicensed and incumbent users of these bands.